%% file: main.tex
\documentclass[submission,copyright,creativecommons]{eptcs}
\providecommand{\event}{LFMTP 2015}

\input{def/packages}
\input{def-import/macros}
\input{def-import/medsquare}
\input{def-import/cong}
\input{def-import/MnSymbol-vdots}
\input{def-import/column-types}
\input{def-import/theorem-envs}

\input{def-import/line-shift}
\input{def/layout-tweaks}

\input{def/macros}

\def\titleText{Proof-relevant \piCalculus}
\title{\titleText}
\def\titlerunning{\titleText}

\input{authors}

\begin{document}
\maketitle

\input{abstract}
\input{sec/introduction}
\input{sec/calculus}

\input{sec/concurrent-transitions}
\input{sec/causal-equivalence}
\input{sec/related-work}
\input{sec/conclusion}
\input{acknowledgements}

\bibliographystyle{eptcs}
\bibliography{../../bib}

\pagebreak
\input{sec/appendix}

\end{document}

%% file: def/packages.tex
\edef\keptrmdefault{\rmdefault}
\edef\keptsfdefault{\sfdefault}
\edef\keptttdefault{\ttdefault}
\usepackage[math]{iwona}
\edef\rmdefault{\keptrmdefault}
\edef\sfdefault{\keptsfdefault}
\edef\ttdefault{\keptttdefault}

\usepackage{float}
\usepackage{mathpartir}
\usepackage{amsmath}
\usepackage{amssymb}
\usepackage{amsthm}
\usepackage{stmaryrd}
\usepackage{graphicx}
\usepackage{soul}
\usepackage[usenames,dvipsnames]{color}
\usepackage{xspace}
\usepackage{mathtools}
\usepackage{url}
\usepackage{hyperref}
\usepackage{breakurl}
\usepackage[T1]{fontenc}
\usepackage{libertine}
\usepackage[scaled=0.75]{beramono}
\usepackage{upgreek}
\usepackage{pifont}
\usepackage{framed}
\usepackage{tikz}
\usepackage{dashundergaps}
\usepackage{scalerel}
\usepackage{array}
\usepackage{bbold}
\usepackage{subcaption}
\usepackage{mathdots}

\usetikzlibrary{matrix} 

\makeatletter
\let\fontaxes@warn@undecodable\@gobble
\makeatother

%% file: def-import/macros.tex
\newcommand{\etal}{et al\xspace}

\newcommand{\eqdef}{\stackrel{\smash{\text{\tiny def}}}{=}}

\DeclareSymbolFont{bbsymbol}{U}{bbold}{m}{n}
\DeclareMathSymbol{\bbsemi}{\mathbin}{bbsymbol}{"3B}

\newcommand{\lowlight}[1]{\textcolor{darkgray}{#1}}
\newcommand{\sub}[2]{#1_{\lowlight{#2}}}
\renewcommand{\sup}[2]{#1^{\lowlight{#2}}}

\renewcommand{\vdots}{\MnSymbolvdots}

\DeclareFontFamily{U}{MnSymbolD}{}
\DeclareFontShape{U}{MnSymbolD}{m}{n}{
    <-6>  MnSymbolD5
   <6-7>  MnSymbolD6
   <7-8>  MnSymbolD7
   <8-9>  MnSymbolD8
   <9-10> MnSymbolD9
  <10-12> MnSymbolD10
  <12->   MnSymbolD12}{}
\DeclareFontShape{U}{MnSymbolD}{b}{n}{
    <-6>  MnSymbolD-Bold5
   <6-7>  MnSymbolD-Bold6
   <7-8>  MnSymbolD-Bold7
   <8-9>  MnSymbolD-Bold8
   <9-10> MnSymbolD-Bold9
  <10-12> MnSymbolD-Bold10
  <12->   MnSymbolD-Bold12}{}
\DeclareSymbolFont{MnSyD}{U}{MnSymbolD}{m}{n}

\DeclareMathSymbol{\sqlt}{\mathrel}{MnSyD}{88}
\DeclareMathSymbol{\sqgt}{\mathrel}{MnSyD}{89}
\DeclareMathSymbol{\sqleq}{\mathrel}{MnSyD}{90}
\DeclareMathSymbol{\sqgeq}{\mathrel}{MnSyD}{91}
\DeclareMathSymbol{\nsqleq}{\mathrel}{MnSyD}{210}
\DeclareMathSymbol{\nsqgeq}{\mathrel}{MnSyD}{211}

\newcommand{\appref}[1]{Appendix~\ref{app:#1}}

\newcommand{\defref}[1]{Definition~\ref{def:#1}}

\newcommand{\figref}[1]{Figure~\ref{fig:#1}}

\newcommand{\lemref}[1]{Lemma~\ref{lem:#1}}

\newcommand{\lemrefThree}[3]{Lemmas \ref{lem:#1}, \ref{lem:#2} and \ref{lem:#3}}

\newcommand{\secref}[1]{Section~\ref{sec:#1}}

\newcommand{\Secref}[1]{\S\ref{sec:#1}}
\newcommand{\thmref}[1]{Theorem~\ref{thm:#1}}
\newcommand{\thmrefTwo}[2]{Theorems~\ref{thm:#1} and \ref{thm:#2}}

\newenvironment{nop}{}{}
\newenvironment{sdisplaymath}{
\begin{nop}\small\begin{displaymath}}{
\end{displaymath}\end{nop}\ignorespacesafterend}
\newenvironment{smathpar}{
\begin{nop}\small\begin{mathpar}}{
\end{mathpar}\end{nop}\ignorespacesafterend}

\newenvironment{mathfig}{\begin{sdisplaymath}}{\end{sdisplaymath}}
\newenvironment{syntaxfig}{\begin{mathfig}\begin{array}{@{}l@{\quad}r@{~~}c@{\quad}ll}}{\end{array}\end{mathfig}}

\makeatletter
\newbox\sf@box
  {\def\sf@one{#1}%
   \def\sf@two{#2}%
   \setbox\sf@box\hbox
     \bgroup}%
  { \egroup
   \ifx\@empty\sf@two\@empty\relax
     \def\sf@two{\@empty}
   \fi
   \ifx\@empty\sf@one\@empty\relax
     \subfloat[\sf@two]{\box\sf@box}%
   \else
     \subfloat[\sf@one][\sf@two]{\box\sf@box}%
   \fi}
\makeatother

\definecolor{highlightcolor}{rgb}{1.0,0.8,0.8}
\definecolor{shadecolor}{rgb}{0.9,0.9,0.9}
\definecolor{lightgray}{rgb}{0.8,0.8,0.8}
\newcommand{\shadebox}[1]{\fcolorbox{lightgray}{shadecolor}{\strut{#1}}}

\DeclareFontFamily{U}{matha}{\hyphenchar\font45}
\DeclareFontShape{U}{matha}{m}{n}{
      <5> <6> <7> <8> <9> <10> gen * matha
      <10.95> matha10 <12> <14.4> <17.28> <20.74> <24.88> matha12
      }{}
\DeclareSymbolFont{matha}{U}{matha}{m}{n}

\DeclareMathSymbol{\twoPrime}{3}{matha}{"32}
\DeclareMathSymbol{\threePrime}{3}{matha}{"33}
\DeclareMathSymbol{\fourPrime}{3}{matha}{"34}

\newenvironment{nscenter}
 {\parskip=0pt\par\nopagebreak\centering}
 {\par\noindent\ignorespacesafterend}

%% file: def-import/medsquare.tex
\DeclareFontFamily{U}{MnSymbolC}{}
\DeclareFontShape{U}{MnSymbolC}{m}{n}{
    <-6>  MnSymbolC5
   <6-7>  MnSymbolC6
   <7-8>  MnSymbolC7
   <8-9>  MnSymbolC8
   <9-10> MnSymbolC9
  <10-12> MnSymbolC10
  <12->   MnSymbolC12}{}
\DeclareFontShape{U}{MnSymbolC}{b}{n}{
    <-6>  MnSymbolC-Bold5
   <6-7>  MnSymbolC-Bold6
   <7-8>  MnSymbolC-Bold7
   <8-9>  MnSymbolC-Bold8
   <9-10> MnSymbolC-Bold9
  <10-12> MnSymbolC-Bold10
  <12->   MnSymbolC-Bold12}{}
\DeclareSymbolFont{MnSyC}{U}{MnSymbolC}{m}{n}

\DeclareMathSymbol{\medsquare}{\mathord}{MnSyC}{106}

%% file: def-import/cong.tex
\DeclareFontFamily{U}{MnSymbolD}{}
\DeclareFontShape{U}{MnSymbolD}{m}{n}{
    <-6>  MnSymbolD5
   <6-7>  MnSymbolD6
   <7-8>  MnSymbolD7
   <8-9>  MnSymbolD8
   <9-10> MnSymbolD9
  <10-12> MnSymbolD10
  <12->   MnSymbolD12}{}
\DeclareFontShape{U}{MnSymbolD}{b}{n}{
    <-6>  MnSymbolD-Bold5
   <6-7>  MnSymbolD-Bold6
   <7-8>  MnSymbolD-Bold7
   <8-9>  MnSymbolD-Bold8
   <9-10> MnSymbolD-Bold9
  <10-12> MnSymbolD-Bold10
  <12->   MnSymbolD-Bold12}{}
\DeclareSymbolFont{MnSyD}{U}{MnSymbolD}{m}{n}

\DeclareMathSymbol{\Cong}{\mathrel}{MnSyD}{12}

%% file: def-import/MnSymbol-vdots.tex
\DeclareFontFamily{U}{MnSymbolC}{}
\DeclareFontShape{U}{MnSymbolC}{m}{n}{
    <-6>  MnSymbolC5
   <6-7>  MnSymbolC6
   <7-8>  MnSymbolC7
   <8-9>  MnSymbolC8
   <9-10> MnSymbolC9
  <10-12> MnSymbolC10
  <12->   MnSymbolC12}{}
\DeclareFontShape{U}{MnSymbolC}{b}{n}{
    <-6>  MnSymbolC-Bold5
   <6-7>  MnSymbolC-Bold6
   <7-8>  MnSymbolC-Bold7
   <8-9>  MnSymbolC-Bold8
   <9-10> MnSymbolC-Bold9
  <10-12> MnSymbolC-Bold10
  <12->   MnSymbolC-Bold12}{}
\DeclareSymbolFont{MnSyC}{U}{MnSymbolC}{m}{n}

\DeclareMathSymbol{\MnSymbolvdots}{\mathord}{MnSyC}{6}

%% file: def-import/column-types.tex
\newcolumntype{L}[1]{>{\raggedright\let\newline\\\arraybackslash\hspace{0pt}}p{#1}}

%% file: def-import/theorem-envs.tex
\newtheorem{theorem}{Theorem}

\newtheorem{lemma}{Lemma}
\theoremstyle{definition}
\newtheorem{definition}{Definition}

%% file: def-import/line-shift.tex
\usetikzlibrary{decorations}
\pgfdeclaredecoration{simple line}{initial}{
  \state{initial}[width=\pgfdecoratedpathlength-1sp]{\pgfmoveto{\pgfpointorigin}}
  \state{final}{\pgflineto{\pgfpointorigin}}
}
\tikzset{
   shift left/.style={decorate,decoration={simple line,raise=#1}},
   shift right/.style={decorate,decoration={simple line,raise=-1*#1}},
}

%% file: def/layout-tweaks.tex
\everymath{\displaystyle}

\allowdisplaybreaks

\usepackage{microtype}

\hypersetup{ 
    colorlinks=false,
    pdfborder={0 0 0},
}

\newcommand{\crossrule}{\noindent\textcolor{lightgray}{\cleaders\hbox{.}\hfill}}
\newcommand{\smathparscale}{0.9}
\newcommand{\Paragraph}[1]{\textbf{#1}.}

\newcommand{\zerodisplayskips}{%
  \setlength{\abovedisplayshortskip}{0pt}
  \setlength{\belowdisplayshortskip}{0pt}}

\appto{\small}{\zerodisplayskips}

%% file: def/macros.tex
\newcommand{\param}{\cdot}
\newcommand{\ruleName}[1]{\textnormal{\textcolor{blue}{#1}}}

\newcommand{\piBoundOutput}[1]{\compl{#1}}

\newcommand{\piInput}[1]{\underline{#1}}
\newcommand{\piOutput}[2]{\compl{#1}\langle#2\rangle}
\newcommand{\piTau}{\tau}

\newcommand{\piAction}[2]{#1.#2}
\newcommand{\piChoice}[2]{{#1} + {#2}}
\newcommand{\piChoiceL}[2]{{#1} + {#2}}
\newcommand{\piChoiceR}[2]{{#1} + {#2}}
\newcommand{\piPar}[2]{{#1} \mid {#2}}
\newcommand{\piParL}[3]{{#2} \mathbin{\lowlight{^{#1}}{\mid}} {#3}}
\newcommand{\piParR}[3]{{#2} \mathbin{\sup{\mid}{#1}} {#3}}
\newcommand{\piParLNu}[2]{{#1} \mathbin{\sub{\sup{\mid}{\tau}}{\nu}} {#2}}
\newcommand{\piParRNu}[2]{{#1} \mathbin{\lowlight{^{\tau}_{\nu}}{\mid}} {#2}}
\newcommand{\piParLTau}[3]{{#1} \mathbin{\sub{\sup{\mid}{\tau}}{#3}} {#2}}
\newcommand{\piParRTau}[3]{{#1} \mathbin{\lowlight{^{\tau}_{#3}}{\mid}} {#2}}
\newcommand{\piReplicate}[1]{{!#1}}
\newcommand{\piRestrictN}[2]{(\nu#1)\;#2}
\newcommand{\piRestrict}[1]{\nu#1}
\newcommand{\piRestrictA}[2]{\sup{\nu}{#1}#2}
\newcommand{\piRestrictOutput}[1]{\compl{\nu}#1}
\newcommand{\piZero}[0]{\mathbf{0}}

\newcommand{\sourceF}{\textsf{src}}
\newcommand{\targetF}{\textsf{tgt}}
\newcommand{\source}[1]{\sourceF({#1})}
\newcommand{\target}[1]{\targetF({#1})}

\newlength{\arrowlen}
\newcommand*{\myrightarrow}[1]{\xrightarrow{\mathmakebox[\arrowlen]{#1}}}

\newcommand{\transition}[1]{\myrightarrow{\smash{#1}}}
\newcommand{\transitionWithoutSmash}[1]{\myrightarrow{#1}}

\newcommand{\compl}[1]{\overline{#1}}

\newcommand{\cxtRaw}[2]{\sup{#1}{#2}}

\newcommand{\concur}{\smile}
\newcommand{\residual}[2]{{#1}\slash{#2}}

\newcommand{\Proc}[1]{\textsf{Proc}\;{#1}}
\newcommand{\action}[1]{\textsf{Action}\;{#1}}

\newcommand{\actions}[1]{\textsf{Action}^*\;{#1}}

\newcommand{\plus}[2]{#1 + #2}
\newcommand{\suc}[1]{\plus{#1}{1}}
\newcommand{\ren}[2]{\renRaw{#1} {#2}}
\newcommand{\renRaw}[1]{{#1}^*}

\newcommand{\id}{\textsf{id}}
\newcommand{\swap}[1]{\sub{\swapR}{#1}}
\newcommand{\swapR}{\textsf{swap}}
\newcommand{\pop}[2]{\sub{\popR}{{#1}}\;{#2}}
\newcommand{\popR}{\textsf{pop}}
\newcommand{\push}[1]{\sub{\pushR}{#1}}
\newcommand{\pushR}{\textsf{push}}

\newcommand{\congEq}{\Cong}
\newcommand{\congRestrictSwap}[1]{\sub{\nu\nu\text{-}\swapR}{#1}}
\newcommand{\congRestrictSwapInv}[1]{\sub{\nu\nu\text{-}\swapR^{-1}}{#1}}
\newcommand{\congTransOp}{\circ}
\newcommand{\congTrans}[2]{{#1} \congTransOp {#2}}
\newcommand{\congRefl}[1]{{#1}_{\congEq}}

\newcommand{\permEq}{\simeq}

\newcommand{\permEqTrans}[2]{{#1} \circ {#2}}
\newcommand{\permEqSwap}[3]{\cons{({#1} \mathbin{:\rightleftharpoons:} {#2})}{#3}}

\newcommand{\cons}[2]{#1\mathrel{::}#2}
\newcommand{\nil}{[]}
\newcommand{\trEmpty}{\nil}
\newcommand{\trCons}[2]{\cons{#1}{#2}}
\newcommand{\asEmpty}{\nil}
\newcommand{\asCons}[2]{\cons{#1}{#2}}

\newcommand{\cofinNew}[1]{\mathrel{\sub{\Join}{#1}}}
\newcommand{\cofinVert}[2]{\mathrel{\sub{\sup{\Join}{#1}}{#2}}}
\newcommand{\rewire}[2]{\sub{\textsf{cofin}}{{#1},{#2}}}
\newcommand{\rewireVec}[1]{\sub{\textsf{cofin}}{{#1}}}
\newcommand{\braid}[1]{\textsf{braid}_{#1}}

\renewcommand{\vec}[1]{\widetilde{#1}}
\newcommand{\after}{\circ}

\renewcommand{\to}{\longrightarrow}

\newcommand{\piCalculus}{$\pi$-calculus\xspace}

\renewcommand{\secref}[1]{\Secref{#1}} 

%% file: authors.tex
\author{Roly Perera
\institute{University of Glasgow\\
Glasgow, UK}
\email{roly.perera@glasgow.ac.uk}
\and
James Cheney
\institute{University of Edinburgh\\
Edinburgh, UK}
\email{jcheney@inf.ed.ac.uk}
}

\def\authorrunning{R. Perera and J. Cheney}

%% file: abstract.tex
\begin{abstract}
  Formalising the \piCalculus is an illuminating test of the
  expressiveness of logical frameworks and mechanised metatheory
  systems, because of the presence of name binding, labelled transitions
  with name extrusion, bisimulation, and structural congruence.
  Formalisations have been undertaken in a variety of systems, primarily
  focusing on well-studied (and challenging) properties such as the
  theory of process bisimulation. We present a formalisation in Agda
  that instead explores the theory of concurrent transitions,
  residuation, and causal equivalence of traces, which has not
  previously been formalised for the \piCalculus. Our formalisation
  employs de Bruijn indices and dependently-typed syntax, and aligns the
  ``proved transitions'' proposed by Boudol and Castellani in the
  context of CCS with the proof terms naturally present in Agda's
  representation of the labelled transition relation. Our main
  contributions are proofs of the ``diamond lemma'' for residuation of
  concurrent transitions and a formal definition of equivalence of
  traces up to permutation of transitions.
\end{abstract}

%% file: sec/introduction.tex
\section{Introduction}

The \piCalculus~\cite{milner99,milner92} is an expressive model of
concurrent and mobile processes.  It has been investigated extensively
and many variations, extensions and refinements have been
proposed, including the asynchronous, polyadic, and applied
\piCalculus (among many others).
The \piCalculus has also attracted considerable attention from the
logical frameworks and meta-languages community, and formalisations of
its syntax and semantics have been performed using most of the extant
mechanised metatheory techniques, including (among others)
Coq~\cite{hirschkoff97,despeyroux00,honsell01},
Nominal Isabelle~\cite{bengtson09b}, Abella~\cite{baelde14} (building
on Miller and Tiu~\cite{tiu10}),
CLF~\cite{cervesato02}, and Agda~\cite{orchard15}. These formalisations
have overcome challenges that tested the limits of these systems (at
least at the time), particularly relating to the encoding of name
binding, scope extrusion and structural congruence. Indeed, some early
formalisations motivated or led to important contributions to the
understanding of these issues in different systems, such as the Theory
of Contexts, or CLF's support for monadic encapsulation of concurrent
executions.

Prior formalisations have typically considered the syntax, semantics
(usually via a variation on labelled transitions), and bisimulation
theory of the \piCalculus. However, as indicated above, while these
aspects of the \piCalculus are essential, they only scratch the surface
of the properties that could be investigated. Most of these developments
have been carried out using informal paper proofs, and formalising them
may reveal challenges or motivate further research on logical
frameworks.

One interesting aspect of the \piCalculus that has not been formally
investigated, and remains to some extent ill-understood informally, is
its theory of \emph{causal equivalence}. Two transitions $t_1,t_2$ that
can be taken from a process term $p$ are said to be \emph{concurrent}
($t_1 \concur t_2$) if they could be performed ``in either order'' ---
that is, if after performing $t_1$, there is a natural way to transform
the other transition $t_2$ so that its effect is performed on the result
of $t_1$, and vice versa. The translation of the second transition is
said to be the \emph{residual} of $t_2$ after $t_1$, written $t_2/t_1$.
The key property of this operation, called the ``diamond lemma'', is
that the two residuals $t_1/t_2$ and $t_2/t_1$ of transitions $t_1
\concur t_2$ result in the same process. Finally, permutation of
concurrent transitions induces a \emph{causal equivalence} relation on
pairs of traces. This is the standard notion of permutation-equivalence
from the theory of traces over concurrent
alphabets~\cite{mazurkiewicz87}.

Our interest in this area stems from previous work on provenance,
slicing and explanation (e.g.~\cite{perera12a}), which we wish to
adapt to concurrent settings. Ultimately, we would like to formalise
the relationship between informal ``provenance graphs'' often used
informally to represent causal relationships~\cite{cheney14} and the
semantics of concurrent languages and traces.  The \piCalculus is a
natural starting point for this study. We wish to understand how to
represent, manipulate, and reason about \piCalculus execution traces
safely: that is, respecting well-formedness and causality.

In classical treatments, starting with L\'evy~\cite{lévy80}, a
transition is usually considered to be a triple $(e,t,e')$ where $e$ and
$e'$ are the terms and $t$ is some information about the step performed.
Boudol and Castellani~\cite{boudol89} introduced the \emph{proved
  transitions} approach for CCS in which the labels of transitions are
enriched with more information about the transition performed. Boreale
and Sangiorgi~\cite{boreale98} and Degano and Priami~\cite{degano99}
developed theories of causal equivalence for the \piCalculus, building
indirectly on the proved transition approach; Danos and
Krivine~\cite{danos04a} and Cristescu, Krivine and
Varacca~\cite{cristescu13} developed notions of causality in the context
of reversible CCS and \piCalculus respectively. However, there does not
appear to be a consensus about the correct definition of causal
equivalence for the \piCalculus. For example, Cristescu
\etal.~\cite{cristescu13} write ``[in] the absence of an indisputable
definition of permutation equivalence for [labelled transition system]
semantics of the \piCalculus it is hard to assert the correctness of one
definition over another.'' In their work on reversible \piCalculus, they
noted that some previous treatments of causality in the \piCalculus did
not allow permuting transitions within the scope of a $\nu$-binder, and
showed how their approach would allow this. Moreover, none of the above
approaches has been formalised.

In this paper, we report on a new formalisation of the \piCalculus
carried out in the dependently-typed programming language
Agda~\cite{norell09}. Our main contributions include formalisations of
concurrency, residuation, the diamond lemma, and causal equivalence. We
do not attempt to formalise the above approaches directly, any one of
which seems to be a formidable challenge. Instead, we have chosen to
adapt the ideas of Boudol and Castellani to the \piCalculus as directly
as we can, guided by the hypothesis that their notion of \emph{proved
  transitions} can be aligned with the \emph{proof terms} for transition
steps that arise naturally in a constructive setting. For example, we
define the concurrency relation on (compatibly-typed) transition proof
terms, and we define residuation as a total function taking two
transitions along with a proof that the transitions are concurrent,
rather than having to deal with a partial operation.

Our formalisation employs de Bruijn indices~\cite{debruijn72}, an
approach with well-known strengths and weaknesses compared, for
example, to higher-order or nominal abstract syntax techniques
employed in existing formalisations.
For convenience, we employ a restricted form of structural congruence
called \emph{braiding congruence}, and we have not formalised as many of
the classical results on the \piCalculus as others have, but we do not
believe there are major obstacles to filling these gaps. To the best of
our knowledge, ours is the first mechanised proof of the diamond lemma
for any process calculus.

The rest of the paper is organised as follows. \secref{calculus}
presents our variant of the (synchronous) \piCalculus, including syntax,
renamings, transitions and braiding congruence.
\secref{concurrent-transitions} presents our definitions of concurrency
and residuation for transitions, and discusses the diamond lemma.
\secref{causal-equivalence} presents our definition of causal
equivalence. \secref{related-work} discusses related work in greater
detail and \secref{conclusion} concludes and discusses prospects for
future work. \appref{module-structure} summarises the Agda module
structure; the source code can be found at
\url{https://github.com/rolyp/proof-relevant-pi}, release $0.1$.
\appref{renaming-lemmas} contains graphical proof-sketches for some
lemmas, and \appref{additional-proofs} some further examples of
residuation.

%% file: sec/calculus.tex
\section{Synchronous \piCalculus}
\label{sec:calculus}

\input{sec/calculus/syntax.tex}
\input{sec/calculus/renamings.tex}
\input{sec/calculus/transitions.tex}

%% file: sec/calculus/syntax.tex

We present our formalisation in the setting of a first-order,
synchronous, monadic \piCalculus with recursion and internal choice,
using a labelled transition semantics. The syntax of the calculus is
conventional (using de Bruijn indices) and is given below.

\vspace{-5pt}
\input{fig/syntax/basic}

Names are ranged over by $x$, $y$ and $z$. An input action is written
$\piInput{x}$. Output actions are written $\piOutput{x}{y}$ if $y$ is in
scope and $\piBoundOutput{x}$ if the action represents the output of a
name whose scope is extruding, in which case we say the action is a
\emph{bound} output. Bound outputs do not appear in user code but arise
during execution.

To illustrate, the conventional $\pi$-calculus term
$\piRestrictN{x}{x(z).\compl{y}\langle{z}\rangle.0 \mid \compl{x}\langle
  c \rangle.0}$ would be represented using de Bruijn indices as $\nu
(\piInput{0}.\piOutput{\suc{n}}{0}.\piZero\mid
\piOutput{0}{\suc{m}}.\piZero)$, provided that $y$ and $c$ are
associated with indices $n$ and $m$. Here, the first $0$ represents the
bound variable $x$, the second $0$ the bound variable $z$, and the third
refers to $x$ again. Note that the symbol $\piZero$ denotes the inactive
process term, not a de Bruijn index.

Let $\Gamma$ and $\Delta$ range over \emph{contexts}, which are finite
initial segments of the natural numbers. The function which extends a
context with a new element is written as a postfix $\suc{\param}$. A
context $\Gamma$ \emph{closes} $P$ if $\Gamma$ contains the free
variables of $P$. We denote by $\Proc{\Gamma}$ the set of processes
closed by $\Gamma$, as defined below. We write $\Gamma \vdash P$ to mean
$P \in \Proc{\Gamma}$. Similarly, actions are well-formed only in closing contexts; we write
$a: \action{\Gamma}$ to mean that $\Gamma$ is closing for $a$, as
defined below.

\vspace{4pt}
\input{fig/syntax/process}
\input{fig/syntax/action}
\vspace{3pt}

To specify the labelled transition semantics, it is convenient to
distinguish \emph{bound} actions $b$ from non-bound actions $c$. A bound
action $b: \action{\Gamma}$ is of the form $\piInput{x}$ or
$\piBoundOutput{x}$, and shifts a process from $\Gamma$ to a target
context $\suc{\Gamma}$, freeing the index $0$. A non-bound action $c:
\action{\Gamma}$ is of the form $\piOutput{x}{y}$ or $\piTau$, and has a
target context which is also $\Gamma$. Meta-variable $a$ ranges over all
actions, bound and non-bound.

%% file: fig/syntax/basic.tex
\begin{minipage}[t]{0.45\linewidth}
\begin{syntaxfig}
\mbox{Name}
&
x, y, z & ::= & 0 \mid 1 \mid \cdots
\\[1mm]
\mbox{Action}
&
a
&
::=
&
\piInput{x}
&
\quad\text{input}
\\
&&&
\piOutput{x}{y}
&
\quad\text{output}
\\
&&&
\piBoundOutput{x}
&
\quad\text{bound output}
\\
&&&
\piTau
&
\quad\text{silent}
\end{syntaxfig}
\end{minipage}%
\begin{minipage}[t]{0.5\linewidth}
\begin{syntaxfig}
\mbox{Process}
&
P, Q, R, S
&
::=
&
\piZero
&
\quad\text{inactive}
\\
&&&
\piAction{\piInput{x}}{P}
&
\quad\text{input}
\\
&&&
\piAction{\piOutput{x}{y}}{P}
&
\quad\text{output}
\\
&&&
\piChoice{P}{Q}
&
\quad\text{choice}
\\
&&&
\piPar{P}{Q}
&
\quad\text{parallel}
\\
&&&
\piRestrict{P}
&
\quad\text{restriction}
\\
&&&
\piReplicate{P}
&
\quad\text{replication}
\end{syntaxfig}
\end{minipage}

%% file: fig/syntax/process.tex
\noindent \shadebox{$\Gamma \vdash P$}
\begin{smathpar}
\inferrule*
{
}
{
  \Gamma \vdash \piZero
}
\and
\inferrule*[right={$x \in \Gamma$}]
{
  \suc{\Gamma} \vdash P
}
{
  \Gamma \vdash \piAction{\piInput{x}}{P}
}
\and
\inferrule*[right={$x, y \in \Gamma$}]
{
  \Gamma \vdash P
}
{
  \Gamma \vdash \piAction{\piOutput{x}{y}}{P}
}
\and
\inferrule*
{
  \Gamma \vdash P
  \\
  \Gamma \vdash Q
}
{
  \Gamma \vdash \piChoice{P}{Q}
}
\and
\inferrule*
{
  \Gamma \vdash P
  \\
  \Gamma \vdash Q
}
{
  \Gamma \vdash \piPar{P}{Q}
}
\and
\inferrule*
{
  \suc{\Gamma} \vdash P
}
{
  \Gamma \vdash \piRestrict{P}
}
\and
\inferrule*
{
  \Gamma \vdash P
}
{
  \Gamma \vdash \piReplicate{P}
}
\end{smathpar}

%% file: fig/syntax/action.tex
\noindent \shadebox{$a : \action{\Gamma}$}
\begin{smathpar}
\inferrule*[right={$x \in \Gamma$}]
{
  \strut
}
{
  \piInput{x}: \action{\Gamma}
}
\and
\inferrule*[right={$x \in \Gamma$}]
{
  \strut
}
{
  \piBoundOutput{x}: \action{\Gamma}
}
\and
\inferrule*[right={$x, y \in \Gamma$}]
{
  \strut
}
{
  \piOutput{x}{y}: \action{\Gamma}
}
\and
\inferrule*
{
  \strut
}
{
  \piTau: \action{\Gamma}
}
\end{smathpar}

%% file: sec/calculus/renamings.tex
\subsection{Renamings}

A de Bruijn indices formulation of \piCalculus makes extensive use of
renamings. A \emph{renaming} $\rho : \Gamma \to \Delta$ is any function
(injective or otherwise) from $\Gamma$ to $\Delta$. The labelled
transition semantics makes use of the lifting of the successor function
$\suc{\param}$ on natural numbers to renamings, which we call $\push{}$
to avoid confusion with the $\suc{\param}$ operation on contexts;
$\pop{}{y}$ which undoes the effect of $\push{}$, replacing $0$ by $y$;
and $\swapR$, which transposes the roles of $0$ and $1$. This de Bruijn
treatment of \piCalculus is similar to that of Hirschkoff's asynchronous
$\mu s$ calculus \cite{hirschkoff99}, except that we give a late rather
than early semantics; other differences are discussed in
\secref{related-work} below.

\vspace{-10pt}
\input{fig/renaming/push-pop-swap}

\noindent The $\Gamma$ subscripts that appear on $\push{\Gamma}$,
$\pop{\Gamma}{x}$ and $\swap{\Gamma}$ are shown in grey to indicate that
they may be omitted when their value is obvious or irrelevant; this is a
convention we use throughout the paper.

\subsubsection{Lifting renamings to processes and actions}
\label{sec:calculus:renamings:processes-and-actions}

The functorial extension $\renRaw{\rho}: \Proc{\Gamma} \longrightarrow
\Proc{\Delta}$ of a renaming $\rho: \Gamma \longrightarrow \Delta$ to
processes is defined in the usual way. Renaming under a binder utilises
the action of $\suc{\param}$ on renamings, which is also functorial.
Syntactically, $\renRaw{\rho}$ binds tighter than any process
constructor.

\begin{minipage}[t]{0.49\linewidth}
\input{fig/renaming/process}
\end{minipage}%
\begin{minipage}[t]{0.49\linewidth}
\input{fig/renaming/action}
\input{fig/renaming/suc}
\end{minipage}

\subsubsection{Properties of renamings}
\label{sec:calculus:properties-of-renamings}

Several equational properties of renamings are used throughout the
development; here we present the ones mentioned elsewhere in the paper.
Diagrammatic versions of the lemmas, along with string diagrams that
offer a graphical intuition for why the lemmas hold, are given in
\appref{renaming-lemmas}.

\begin{lemma}
\label{lem:pop-after-push}
$\pop{}{x} \after \push{} = \id$
\end{lemma}

\noindent Freeing the index $0$ and then immediately
substituting $x$ for it is a no-op.

\begin{lemma}
\label{lem:pop-zero-after-suc-push}
$\pop{}{0} \after \suc{\push{}} = \id$
\end{lemma}

\begin{lemma}
\label{lem:swap-after-suc-swap-after-swap}
$\suc{\swap{}} \after \swap{} \after \suc{\swap{}} = \swap{} \after
\suc{\swap{}} \after \swap{}$
\end{lemma}

\noindent The above are two equivalent ways of swapping indices 0 and 2.

\begin{lemma}
\label{lem:pop-swap}
$\pop{}{0} \after \swap{} = \pop{}{0}$
\end{lemma}

\begin{lemma}
\label{lem:swap-push}
$\swap{} \after \suc{\push{}} = \push{}$, $\swap{} \after \push{} = \suc{\push{}}$
\end{lemma}

\begin{lemma}
$\push{}\after \rho= \suc{\rho} \after \push{}$
\label{lem:push-comm}
\end{lemma}

\begin{lemma}
$\rho \after \pop{}{x} = \pop{}{\rho x} \after \suc{\rho}$
\label{lem:pop-comm}
\end{lemma}

\begin{lemma}
\label{lem:swap-suc-suc}
$\swap{} \after \plus{\rho}{2} = \plus{\rho}{2} \after \swap{}$
\end{lemma}

\noindent These last two lemmas assert various naturality properties of
$\push{}$, $\pop{}{x}$ and $\swap{}$.

%% file: fig/renaming/push-pop-swap.tex
\begin{minipage}[t]{0.32\linewidth}
{\small
\begin{align*}
\intertext{\shadebox{$\push{\Gamma}: \Gamma \longrightarrow \plus{\Gamma}{1}$}}
\push{}\;x
&=
x + 1
\end{align*}
}
\end{minipage}%
\begin{minipage}[t]{0.32\linewidth}
{\small
\begin{align*}
\intertext{\shadebox{$\sub{\popR}{\Gamma}: \Gamma \longrightarrow \plus{\Gamma}{1} \longrightarrow \Gamma$}}
\pop{}{y}\;0
&=
y
\\
\pop{}{y}\;(x + 1)
&=
x
\end{align*}
}
\end{minipage}%
\begin{minipage}[t]{0.32\linewidth}
{\small
\begin{align*}
\intertext{\shadebox{$\swap{\Gamma}: \plus{\Gamma}{2} \longrightarrow \plus{\Gamma}{2}$}}
\swapR\;0
&=
1
\\
\swapR\;1
&=
0
\\
\swapR\;(x + 2)
&=
x + 2
\end{align*}
}
\end{minipage}

%% file: fig/renaming/process.tex
{\small
\begin{align*}
\intertext{\shadebox{$\ren{\param}{}: (\Gamma \to \Delta) \to \Proc{\Gamma} \to \Proc{\Delta}$}}
\ren{\rho}{\piZero}
&=
\piZero
\\
\ren{\rho}{(\piAction{\piInput{x}}{P})}
&=
\piAction{\piInput{\rho x}}{(\ren{\suc{\rho})}{P}}
\\
\ren{\rho}{(\piAction{\piOutput{x}{y}}{P})}
&=
\piAction{\piOutput{\rho x}{\rho y}}{\ren{\rho}{P}}
\\
\ren{\rho}{(\piChoice{P}{Q})}
&=
\piChoice{\ren{\rho}{P}}{\ren{\rho}{Q}}
\\
\ren{\rho}{(\piPar{P}{Q})}
&=
\piPar{\ren{\rho}{P}}{\ren{\rho}{Q}}
\\
\ren{\rho}{(\piRestrict{P})}
&=
\piRestrict{\ren{(\suc{\rho})}{P}}
\\
\ren{\rho}{(\piReplicate{P})}
&=
\piReplicate{\ren{\rho}{P}}
\end{align*}
}

%% file: fig/renaming/action.tex
{\small
\begin{align*}
\intertext{\shadebox{$\ren{\param}{}: (\Gamma \to \Delta) \to \action{\Gamma} \to \action{\Delta}$}}
\ren{\rho}{\;\piInput{x}}
&=
\piInput{\rho x}
\\
\ren{\rho}{\;\piBoundOutput{x}}
&=
\piBoundOutput{\rho x}
\\
\ren{\rho}{\;\tau}
&=
\tau
\\
\ren{\rho}{\;\piOutput{x}{y}}
&=
\smash{\piOutput{\rho x}{\rho y}}
\end{align*}
}

%% file: fig/renaming/suc.tex
{\small
\begin{align*}
\intertext{\shadebox{$\suc{\param}: (\Gamma \to \Delta) \to \suc{\Gamma} \to \suc{\Delta}$}}
(\suc{\rho})\;0
&=
0
\\
(\suc{\rho})\;(x + 1)
&=
\rho x + 1
\end{align*}}

%% file: sec/calculus/transitions.tex
\subsection{Labelled transition semantics}

An important feature of our semantics is that each transition rule has
an explicit constructor name. This allow derivations to be written in a
compact, expression-like form, similar to the \emph{proven transitions}
used by Boudol and Castellani to define notions of concurrency and
residuation for CCS \cite{boudol89}. However, rather than giving an
additional inductive definition describing the structure of a ``proof''
that $P \transition{a} R$, we simply treat the inductive definition of
$\transition{}$ as a data type. This is a natural approach in a
dependently-typed setting.

The rule names are summarised below, and have been chosen to reflect,
where possible, the structure of the process triggering the rule. The
corresponding relation $P \transition{a} R$ is defined in
\figref{transition}, for any process $\Gamma \vdash P$, any $a:
\action{\Gamma}$ with target $\Delta \in \{\Gamma, \suc{\Gamma}\}$, and
any $\Delta \vdash R$.

\input{fig/syntax/transition}

\input{fig/transition}

The constructor name for each rule is shown to the left of the rule.
There is an argument position, indicated by $\param$, for each premise
of the rule. Note that there are two forms of the transition
constructors $\piParL{a}{\param}{\param}$ and $\piRestrictA{a}{\param}$
distinguished by whether they are indexed by a bound action $b$ or by a
non-bound action $c$. Moreover there are additional (but symmetric)
rules of the form $\piChoice{P}{\param}$, $\piParR{b}{P}{\param}$ and
$\piParR{b}{P}{\param}$ where the sub-transition occurs on the opposite
side of the operator, and similarly $\piParRNu{\param}{\param}$ and
$\piParRTau{\param}{\param}{y}$ rules in which the positions of sender
and receiver are transposed. These are all straightforward variants of
the rules shown, and are omitted from \figref{transition} for brevity.
Meta-variables $E$ and $F$ range over transition derivations; if $E: P
\transition{a} R$ then $\source{E}$ denotes $P$ and $\target{E}$ denotes
$R$.

Although a de Bruijn formulation of pi calculus requires a certain
amount of housekeeping, one pleasing consequence is that the usual
side-conditions associated with the \piCalculus transition rules are
either subsumed by syntactic constraints on actions, or
``operationalised'' using the renamings above. In particular:

\begin{enumerate}
\item The use of $\pushR$ in the \ruleName{$\piParL{b}{\param}{Q}$} rule
  corresponds to the usual side-condition asserting that the binder
  being propagated by $P$ is not free in $Q$. In the de Bruijn setting
  every binder ``locally'' has the name 0, and so this requirement can
  be operationalised by rewiring $Q$ so that the name $0$ is reserved.
  The $\pushR$ will be matched by a later $\popR$ which substitutes for
  $0$, in the event that the action has a successful rendezvous.
\item The \ruleName{$\piRestrictOutput{\param}$} rule requires an
  extrusion to be initiated by an output of the form
  $\piOutput{\suc{x}}{0}$, capturing the usual side-condition that the
  name being extruded \emph{on} is distinct from the name being
  extruded.
\item The rules of the form \ruleName{$\piRestrictA{a}$} require that
  the action being propagated has the form $\ren{\pushR}{a}$, ensuring
  that it contains no uses of index $0$. This corresponds to the usual
  requirement that an action can only propagate through a binder that it
  does not mention.
\end{enumerate}

\noindent The use of $\swapR$ in the \ruleName{$\piRestrictA{b}$} case
follows Hirschkoff \cite{hirschkoff99} and has no counterpart outside of
the de Bruijn setting. As a propagating binder passes through another
binder, their local names are 0 and 1. Propagation transposes the
binders, and so to preserve naming we rewire $R$ with a ``braid'' that
swaps $0$ and $1$. Since binders are also reordered by
\emph{permutations} that relate causally equivalent executions, the
$\swapR$ renaming will also play an important role when we consider
concurrent transitions (\secref{concurrent-transitions}).

The following schematic derivation shows how the compact notation works.
Suppose $E: P \transition{\piOutput{\plus{z}{2}}{0}} R$ takes place
immediately under a $\nu$-binder, causing the scope of the binder to be
extruded. Then suppose the resulting bound output propagates through
another binder, giving the partial derivation on the left:

\input{fig/transition/illustrate/compact-notation}

\noindent with $E$ standing in for the rest of the derivation. The blue
constructors annotating the left-hand side of the derivation tree can be
thought of as a partially unrolled ``transition term'' representing the
proof. The $\param$ placeholders associated with each constructor are
conceptually filled by the transition terms annotating the premises of
that step. We can ``roll up'' the derivation by a single step, by moving
the premises into their corresponding placeholders, as shown in the
middle figure.

By repeating this process, we can write the whole derivation compactly
as $\piRestrictA{\piBoundOutput{z}}{\piRestrictOutput{E}}$, as shown on
the right. Thus the compact form is simply a flattened transition
derivation: similar to a simply-typed lambda calculus term written as a
conventional expression, in a (Church-style) setting where a term is,
strictly speaking, a typing derivation.

\subsubsection{Residuals of transitions and renamings}

A transition survives any suitably-typed renaming. As alluded to
already, this will be essential to formalising causal equivalence. First
we define the (rather trivial) residual of a renaming $\rho: \Gamma \to
\Delta$ after an action $a: \action{\Gamma}$.

\begin{definition}[Residual of $\rho$ after $a$]
\begin{align*}
\residual{\rho}{b} &\eqdef \suc{\rho} \\ \residual{\rho}{c} &\eqdef \rho
\end{align*}
\end{definition}

\noindent The complementary residual $\residual{a}{\rho}$ is also
defined and is simply the renamed action $\ren{\rho}{a}$ defined earlier
in \secref{calculus:renamings:processes-and-actions}. We use the latter
notation.

\begin{lemma}
\label{lem:concurrent-transitions:renaming:transition}
Suppose $E: P \transition{a} Q$ and $\rho: \Gamma \to \Delta$, where
$\Gamma \vdash P$. Then there exists a transition $\residual{E}{\rho}:
\ren{\rho}{P} \transition{\ren{\rho}{a}} \ren{(\residual{\rho}{a})}{Q}$
such that $\target{\residual{E}{\rho}} = \ren{\residual{\rho}{a}}{Q}$.

{\small
\begin{nscenter}
\input{fig/residual/renaming}
\end{nscenter}
}
\end{lemma}

\noindent The proof is the obvious lifting of a renaming to a
transition, and is given in \appref{additional-proofs}.

We would not expect $\residual{E}{\rho}$ to be derivable for arbitrary
$\rho$ in all extensions of the \piCalculus. In particular, the mismatch
operator $[x\neq y]P$ that steps to $P$ if $x$ and $y$ are distinct
names is only stable under injective renamings.

\subsubsection{Structural congruences}

We believe our semantics to be closed under the usual \piCalculus
congruences, but have not attempted to formalise this. The ``braiding''
congruence $\congEq$ introduced in
\secref{concurrent-transitions:cofinality} is in fact a standard
\piCalculus congruence, which we use to track changes in the relative
position of binders under permutations of traces. This could be
generalised to include more congruences, but at a corresponding cost in
formalisation complexity.

%% file: fig/syntax/transition.tex
\begin{syntaxfig}
\mbox{Transition}
&
E, F
&
::=
&
\piAction{\piInput{x}}{P}
&
\text{input on $x$}
\\
&&&
\piAction{\piOutput{x}{y}}{P}
&
\text{output $y$ on $x$}
\\
&&&
\piChoiceL{E}{Q} \quad \piChoiceR{P}{F}
&
\text{choose left or right branch}
\\
&&&
\piParL{a}{E}{Q} \quad \piParR{a}{P}{F}
&
\text{propagate $a$ through parallel composition on the left or right}
\\
&&&
\piParLTau{E}{F}{y} \quad \piParRTau{E}{F}{y}
&
\text{rendezvous (receiving $y$ on the left or right)}
\\
&&&
\piRestrictOutput{E}
&
\text{initiate name extrusion}
\\
&&&
\piParLNu{E}{F} \quad \piParRNu{E}{F}
&
\text{extrusion rendezvous (receiving $0$ on the left or right)}
\\
&&&
\piRestrictA{a}{E}
&
\text{propagate $a$ through binder}
\\
&&&
\piReplicate{E}
&
\text{replicate}
\end{syntaxfig}

%% file: fig/transition.tex
\begin{figure}[h]
\noindent \shadebox{$P \transition{a} R$}
\begin{smathpar}
\inferrule*[left={\ruleName{$\piAction{\piInput{x}}{P}$}}]
{
}
{
   \piAction{\piInput{x}}{P} \transitionWithoutSmash{\piInput{x}} P
}
\and
\inferrule*[left={\ruleName{$\piAction{\piOutput{x}{y}}{P}$}}]
{
}
{
   \piAction{\piOutput{x}{y}}{P} \transitionWithoutSmash{\piOutput{x}{y}} P
}
\and
\inferrule*[left={\ruleName{$\piChoiceL{\param}{Q}$}}]
{
   P \transitionWithoutSmash{a} R
}
{
   \piChoice{P}{Q} \transitionWithoutSmash{a} R
}
\and
\inferrule*[left={\ruleName{$\piParL{c}{\param}{Q}$}}]
{
   P \transitionWithoutSmash{c} R
}
{
   \piPar{P}{Q} \transitionWithoutSmash{c} \piPar{R}{Q}
}
\and
\inferrule*[left={\ruleName{$\piParL{b}{\param}{Q}$}}]
{
   P \transitionWithoutSmash{b} R
}
{
   \piPar{P}{Q} \transitionWithoutSmash{b} \piPar{R}{\ren{\push{}}{Q}}
}
\and
\inferrule*[left={\ruleName{$\piParLTau{\param}{\param}{y}$}}]
{
   P \transitionWithoutSmash{\piInput{x}} R
   \\
   Q \transitionWithoutSmash{\piOutput{x}{y}} S
}
{
   \piPar{P}{Q} \transitionWithoutSmash{\piTau} \piPar{\ren{(\pop{}{y})}{R}}{S}
}
\and
\inferrule*[left={\ruleName{$\piRestrictOutput{\param}$}}]
{
   P \transitionWithoutSmash{\piOutput{(x + 1)}{0}} R
}
{
   \piRestrict{P} \transitionWithoutSmash{\piBoundOutput{x}} R
}
\and
\inferrule*[left={\ruleName{$\piParLNu{\param}{\param}$}}]
{
   P \transitionWithoutSmash{\piInput{x}} R
   \\
   Q \transitionWithoutSmash{\piBoundOutput{x}} S
}
{
   \piPar{P}{Q} \transitionWithoutSmash{\piTau} \piRestrict{(\piPar{R}{S})}
}
\and
\inferrule*[left={\ruleName{$\piRestrictA{c}{\param}$}}]
{
   P \transitionWithoutSmash{\ren{\push{}}{c}} R
}
{
   \piRestrict{P} \transitionWithoutSmash{c} \piRestrict{R}
}
\and
\inferrule*[left={\ruleName{$\piRestrictA{b}{\param}$}}]
{
   P \transitionWithoutSmash{\ren{\push{}}{b}} R
}
{
   \piRestrict{P} \transitionWithoutSmash{b} \piRestrict{(\ren{\swapR}{R})}
}
\and
\inferrule*[left={\ruleName{$\piReplicate{\param}$}}]
{
   \piPar{P}{\piReplicate{P}} \transitionWithoutSmash{a} R
}
{
   \piReplicate{P} \transitionWithoutSmash{a} R
}
\end{smathpar}
\crossrule
\caption{Labelled transition rules ($\piChoice{P}{\param}$,
  $\piParR{b}{P}{\param}$, $\piParR{c}{P}{\param}$,
  $\piParRNu{\param}{\param}$ and $\piParRTau{\param}{\param}{y}$
  variants omitted)}
\label{fig:transition}
\end{figure}

%% file: fig/transition/illustrate/compact-notation.tex
\begin{center}
\begin{minipage}[b]{0.3\linewidth}
\scalebox{\smathparscale}{
\begin{smathpar}
\inferrule*[left={\ruleName{$\piRestrictA{\piBoundOutput{z}}{\param}$}}]
{
  \inferrule*[left={\ruleName{$\piRestrictOutput{\param}$}}]
  {
    \inferrule*[left={\ruleName{$E$}}]
    {
      \vdots
    }
    {
      P \transitionWithoutSmash{\piOutput{\plus{z}{2}}{0}} R
    }
  }
  {
    \piRestrict{P} \transitionWithoutSmash{\piBoundOutput{\suc{z}}} R
  }
}
{
  \piRestrict{\piRestrict{P}} \transitionWithoutSmash{\piBoundOutput{z}} \piRestrict{R}
}
\end{smathpar}
}
\end{minipage}
\begin{minipage}[b]{0.3\linewidth}
\scalebox{\smathparscale}{
\begin{smathpar}
\inferrule*[left={\ruleName{$\piRestrictA{\piBoundOutput{z}}{\param}$}}]
{
  \inferrule*[left={\ruleName{$\piRestrictOutput{E}$}}]
  {
    \vdots
  }
  {
    \piRestrict{P} \transitionWithoutSmash{\piBoundOutput{\suc{z}}} R
  }
}
{
  \piRestrict{\piRestrict{P}} \transitionWithoutSmash{\piBoundOutput{z}} \piRestrict{R}
}
\end{smathpar}
}
\end{minipage}
\begin{minipage}[b]{0.3\linewidth}
\scalebox{\smathparscale}{
\begin{smathpar}
\inferrule*[left={\ruleName{$\piRestrictA{\piBoundOutput{z}}{\piRestrictOutput{E}}$}}]
{
  \vdots
}
{
  \piRestrict{\piRestrict{P}} \transitionWithoutSmash{\piBoundOutput{z}} \piRestrict{R}
}
\end{smathpar}
}
\end{minipage}
\end{center}

%% file: fig/residual/renaming.tex
\scalebox{0.8}{
\begin{tikzpicture}[node distance=1.5cm, auto]
  \node (P) {$P$};
  \node (Q) [below of=P] {$\ren{\rho}{P}$};
  \node (PPrime) [node distance=3.4cm, right of=P] {$Q$};
  \node (QPrime) [below of=PPrime] {$\ren{(\residual{\rho}{a})}{Q}$};
  \draw[->] (P) to node {$E$} (PPrime);
  \draw[dotted,->] (Q) to node [swap] {$\residual{E}{\rho}$} (QPrime);
  \draw[->] (P) to node [swap] {$\ren{\rho}{}$} (Q);
  \draw[->] (PPrime) to node {$\ren{(\residual{\rho}{a})}{}$} (QPrime);
\end{tikzpicture}
}

%% file: sec/concurrent-transitions.tex
\section{Concurrency and residuals}
\label{sec:concurrent-transitions}

We now use the compact notation for derivations to define a notion of
\emph{concurrency} for transitions with the same source state, following
the work of Boudol and Castellani for CCS \cite{boudol89}. Concurrent
transitions are independent, or causally unordered: they can execute in
either order without significant interference. Permutation of concurrent
transitions induces a congruence on traces, which is the topic of
\secref{causal-equivalence}.

\subsection{Concurrent transitions}

Transitions $P \transition{a} R$ and $Q \transition{a'} S$ are
\emph{coinitial} iff $P = Q$. We now define a symmetric and irreflexive
relation $\concur$ over coinitial transitions. If $E \concur E'$ we say
$E$ and $E'$ are \emph{concurrent}. The relation is defined as the
symmetric closure of the rules given in \figref{concurrent}, again with
trivial variants of the rules omitted. For the transition constructors
of the form \ruleName{$\piParL{a}{\param}{Q}$} and
\ruleName{$\piRestrictA{a}{\param}$} which come in bound and non-bound
variants, we abuse notation a little and write a single $\concur$ rule
quantified over $a$ to mean that there are two separate (but otherwise
identical) cases.

\input{fig/concurrent}

The first rule, $\piParR{a}{P}{F} \concur \piParL{a'}{E}{Q}$, says that
two transitions $E$ and $F$ are concurrent if they take place on
opposite sides of the same parallel composition. The remaining rules
propagate concurrent sub-transitions up through $\nu$, choice, parallel
composition, and replication. Note that there are no rules allowing us
to conclude that a left-choice step is concurrent with a right-choice
step: choices are mutually exclusive. Likewise, there are no rules
allowing us to conclude that an input or output transition is concurrent
with any other transition; since both $E$ and $E'$ are required to be
coinitial, if one of them is an input or output step then they are equal
and hence not concurrent.

The $\piParLTau{E}{F}{y} \concur \piParLTau{E'}{F'}{z}$ rule says that a
rendezvous is concurrent with another rendezvous under the same parallel
composition, as long as the two inputs are concurrent on the left, and
the two outputs are concurrent on the right. The $\piParLTau{E}{F}{y}
\concur \piParRTau{E'}{F'}{z}$ variant is similar, but permits
concurrent input and output on the left, with their rendezvous partners
concurrent on the right. The $\piParLTau{E}{F}{y} \concur
\piParRNu{E'}{F'}$ rule and variants permit a regular rendezvous and an
extrusion-rendezvous to be concurrent.

\subsection{Residuals of concurrent transitions}
\label{sec:concurrent-transitions:residuals}

Intuitively, if $E \concur E'$ then $E$ and $E'$ are ``parallel moves''
in the sense of Curry and Feys \cite{curry58}: if either execution step
is taken, the other remains valid, and if both are taken, one ends up in
(essentially) the same state, regardless of which step is taken first.

However, concurrent transitions are not completely independent: the
location and nature of the redex identified by one transition may change
as a consequence of the earlier transition. This intuition is captured
by the notion of the residual $\residual{E}{E'}$, explored notably by
L\'evy in the lambda calculus \cite{lévy80}, and later considered by
Stark for concurrent transition systems \cite{stark89} and in the
specific setting of CCS by Boudol and Castellani \cite{boudol89}. The
residual specifies how $E$ must be adjusted to take into account the
fact that $E'$ has taken place.

\begin{definition}[Residual]
Suppose $E \concur E'$. Then the \emph{residual} of $E$ after $E'$,
written $\residual{E}{E'}$, is given by the least function satisfying
the equations in \figref{residual}.
\end{definition}

The operator $\residual{\param}{\param}$ has higher precedence than any
transition constructor. The definition makes use of the renaming lemmas
in \secref{calculus:properties-of-renamings}, and is rather tricky;
\appref{additional-proofs:concurrent-transitions:cofinality} gives
several examples which illustrate some of the subtleties that arise in
the \piCalculus setting, in particular relating to name extrusion.

\input{fig/residual}

\subsubsection{Cofinality of residuals}
\label{sec:concurrent-transitions:cofinality}

The idea that one ends up in the same state regardless of whether $E$ or
$E'$ is taken first is called \emph{cofinality}. In CCS, where actions
never involve binders, and in the lambda calculus, where binders do not
move around, cofinality simply means the target states are equivalent.
Things are not quite so simple in late-style \piCalculus, because
binders propagate during execution, as bound actions. Consider the
process $\piPar{\piAction{\piInput{x}}{P}}{\piAction{\piInput{z}}{Q}}$
with two concurrent input actions. Initiating one of the inputs (say
$\piInput{x}$) starts propagating a binder. As this binder passes
through the parallel composition, the transition rules use $\pushR$ to
``reserve'' the free variable 0 in the right half of the process for
potential use by a subsequent $\popR$:

\input{fig/residual/illustrate/inputs-1.tex}

\noindent When the action ($\piInput{\suc{z}}$) is performed, a $\pushR$
on the left leaves the final state with both $0$ and $1$ reserved:

\input{fig/residual/illustrate/inputs-2.tex}

Had these concurrent actions happened in the opposite order, the
$\pushR$ on the left would have been applied first. The final state
would be $\piPar{\ren{(\suc{\push{}})}{P}}{\ren{\push{}}{Q}}$, which is
the image of $\piPar{\ren{\push{}}{P}}{\ren{(\suc{\push{}})}{Q}}$ in the
permutation $\swapR$ which renames $0$ to $1$ and $1$ to $0$. Instead of
the usual cofinality square, the final states are related by a ``braid''
(in the form of a $\swapR$) which permutes the free names:

\input{fig/residual/illustrate/inputs.tex}

\noindent Here $\alpha$ and $\beta$ are equalities obtained from
\lemref{swap-push}.

It is not just the reordering of bound actions which nuances \piCalculus
cofinality. When two $\tau$ actions are reordered, which happen to be
extrusion rendezvous of distinct binders, the resulting binders exchange
positions in the final process. In the standard \piCalculus this would
be subsumed by the congruence $\piRestrictN{xy}{P} \congEq
\piRestrictN{yx}{P}$. In the de Bruijn setting, where adjacent binders
cannot be distinguished, the analogous rule is
$\piRestrict{\piRestrict{P}} \congEq
\piRestrict{\piRestrict{(\ren{\swapR}{P})}}$, which applies a $\swapR$
braid under the two binders.

These two possibilities are subsumed by the following generalised notion
of cofinality. First we define a braiding congruence $\congEq$ just
large enough to permit $\swapR$ under a pair of binders. ``Cofinality''
is then defined using a more general braiding relation which
additionally permits $\swapR$s of free variables. Examples showing
reordered extrusions are given in
\appref{additional-proofs:concurrent-transitions:cofinality}, including
concurrent extrusions of the \emph{same} binder, an interesting case
identified by Cristescu \etal.~\cite{cristescu13}.

\begin{definition}[Braiding congruence]
Inductively define the binary relation $\congEq$ over processes using
the rules given in \figref{congruence}.
\end{definition}

\input{fig/congruence}

In \figref{congruence}, rule names are shown to the left in blue,
permitting a compact term-like notation for $\congEq$ proofs similar to
the convention we introduced earlier for transitions. The process
constructors are overloaded to witness compatibility; transitivity is
denoted by $\congTransOp$. It is easy to see that $\congEq$ is also
reflexive and symmetric, and therefore a congruence. $\congRefl{P}$
denotes the canonical proof that $P \congEq P$.

In what follows $\phi$ and $\psi$ range over braiding congruences;
$\source{\phi}$ and $\target{\phi}$ denote $P$ and $R$, for any $\phi: P
\congEq R$. As with transitions, braiding congruences are stable under
renamings, giving rise to the usual notion of residuation; however
$\residual{\rho}{\phi}$ is always $\rho$. The proof is a straightforward
induction.

\begin{lemma}
\label{lem:concurrent-transitions:renaming-preserves-cong}
For any $\Gamma \vdash P$, suppose $\phi: P \to Q$ and $\rho: \Gamma \to
\Delta$. Then there exists a braiding congruence $\residual{\phi}{\rho}:
\ren{\rho}{P} \to \ren{\rho}{Q}$.

\begin{nscenter}
\input{fig/renaming/congruence.tex}
\end{nscenter}
\end{lemma}

\begin{definition}[Braiding]
\label{def:concurrent-transitions:cofin}
For any $\Delta \in \{0, 1, 2\}$ define the following family of
bijective renamings $\braid{\Gamma,\Delta}: \plus{\Gamma}{\Delta} \to
\plus{\Gamma}{\Delta}$ and symmetric \emph{braiding} relations
$\cofinNew{\Gamma,\Delta}$ over processes in $\plus{\Gamma}{\Delta}$.

\input{fig/braiding}
\end{definition}

Our key soundness result is that residuals of concurrent transitions $E$
and $E'$ are always cofinal up to a braiding of type
$\cofinNew{\Gamma,\Delta}$ where $\Delta \in \{0, 1, 2\}$ is the number
of free variables introduced by $E$ and $\residual{E'}{E}$. Rather than
the usual parallel-moves square on the left, the residuals satisfy
pentagons of the form shown in the centre of
\figref{concurrent-transitions:cofinality}, where $\gamma: Q
\cofinNew{\Gamma,\Delta} Q'$ is a braiding.

\begin{figure}[h]
\begin{minipage}[t]{0.5\linewidth}
\input{fig/residual/illustrate/square}
\end{minipage}%
\raisebox{0.7em}{
\begin{minipage}[t]{0.49\linewidth}
\input{fig/residual/illustrate/pentagon}
\end{minipage}
}
\caption{Cofinality in the style of CCS (left); with explicit braiding
  (right)}
\label{fig:concurrent-transitions:cofinality}
\end{figure}

\noindent Arranging for this to hold by construction introduces a
certain amount of complexity, so we prove cofinality as a separate
theorem.

\begin{theorem}[Cofinality of residuals]
\label{thm:concurrent-transitions:cofinality}
Suppose $E$ and $E'$ are the transitions on the right of
\figref{concurrent-transitions:cofinality}, with $E \concur E'$. Then
there exists $\rewire{E}{E'}: Q \cofinNew{\Delta} Q'$.
\end{theorem}

The notion of concurrency extends into dimensions greater than two.
Following Pratt's higher-dimensional automata \cite{pratt00}, we can
consider a proof $\chi: E \concur E'$ as a surface that represents the
concurrency of $E$ and $E'$ without committing to an order of
occurrence. Every such $\chi: E \concur E'$ has a two-dimensional
residual $\residual{\chi}{E''}$ with respect to a third concurrent
transition $E''$. First we note that concurrent transitions are closed
under renamings.

\begin{lemma}
\label{lem:concurrent-transitions:renaming-preserves-concurrency}
Suppose $\rho: \Gamma \to \Delta$ and $E$, $E'$ are both transitions
from $\Gamma \vdash P$, with $\chi: E \concur E'$. Then there exists
$\residual{\chi}{\rho}: \residual{E}{\rho} \concur \residual{E'}{\rho}$.
\end{lemma}

\begin{proof}
By induction on $\chi$, using
\lemref{concurrent-transitions:renaming:transition}.
\end{proof}

\begin{theorem}[Residuation preserves concurrency]
\label{thm:concurrent-transitions:residuation-preserves-concurrency}
\item
Suppose $\chi: E \concur E'$ with $E \concur E''$ and $E' \concur E''$.
Then there exists $\residual{\chi}{E''}: \residual{E}{E''} \concur
\residual{E'}{E''}$.
\end{theorem}

\begin{proof}
By induction on $\chi$ and inversion on the other two derivations, using
\lemref{concurrent-transitions:renaming-preserves-concurrency}.
\end{proof}

\begin{theorem}
\label{thm:concurrent-transitions:cofinality:cube}
Suppose $\chi: E \concur E'$, with $E' \concur E''$ and $E'' \concur E$.
Then:
\[\residual{(\residual{(\residual{E'}{E''})}{(\residual{E}{E''})})}{\rewire{E''}{E}}
=
\residual{(\residual{E'}{E})}{(\residual{E''}{E})}\]
\end{theorem}

The diagram below illustrates
\thmrefTwo{concurrent-transitions:residuation-preserves-concurrency}{concurrent-transitions:cofinality:cube}
informally. The three faces $\chi$, $\chi'$ and $\chi''$ with $P$ as a
vertex witness the pairwise concurrency of $E$, $E'$ and $E''$.
\thmref{concurrent-transitions:residuation-preserves-concurrency}
ensures that these have opposite faces $\residual{\chi}{E''}$,
$\residual{\chi'}{E}$ and $\residual{\chi''}{E'}$.
\thmref{concurrent-transitions:cofinality:cube} states that, up to a
suitable braiding, there is a unique residual of a one-dimensional
transition after a concurrent two-dimensional one, connecting the faces
$\residual{\chi'}{E}$ and $\residual{\chi''}{E'}$ via the shared edge
$\residual{E''}{\chi}$. Analogous reasoning for $\residual{E}{\chi'}$
and $\residual{E'}{\chi''}$ yields a cubical transition with target
$\mathbf{P'}$.

\input{fig/concurrency-cube}

The bold font for $\mathbf{S_1}$, $\mathbf{S_2}$, $\mathbf{S_3}$ and
$\mathbf{P'}$ indicates that they represent not a unique process but a
permutation group of processes related by braidings. At $\mathbf{P'}$
there are potentially $3! = 6$ variants of the target process, one for
each possible interleaving of $E$, $E'$ and $E''$. The notation
$\residual{\mathbf{E''}}{\mathbf{\chi}}$ is again informal, referring
not to a unique transition but to a permutation group related by
braidings.

%% file: fig/concurrent.tex
\begin{figure}[h]
\shadebox{$E \concur E'$}
\begin{smathpar}
\inferrule*
{
  \strut
}
{
  \piParR{a}{P}{F}
  \concur
  \piParL{a'}{E}{Q}
}
\and
\inferrule*
{
  E \concur E'
}
{
  \piParL{a}{E}{Q}
  \concur
  \piParLTau{E'}{F}{y}
}
\and
\inferrule*
{
  F \concur F'
}
{
  \piParR{a}{P}{F}
  \concur
  \piParLTau{E}{F'}{y}
}
\and
\inferrule*
{
  E \concur E'
}
{
  \piParL{a}{E}{Q}
  \concur
  \piParLNu{E'}{F}
}
\and
\inferrule*
{
  F \concur F'
}
{
  \piParR{a}{P}{F}
  \concur
  \piParLNu{E}{F'}
}
\and
\inferrule*
{
  E \concur E'
}
{
  \piChoice{E}{Q}
  \concur
  \piChoiceL{E'}{Q}
}
\and
\inferrule*
{
  F \concur F'
}
{
  \piParR{a}{P}{E}
  \concur
  \piParR{a'}{P}{E'}
}
\and
\inferrule*
{
  E \concur E'
}
{
  \piParL{a}{E}{Q}
  \concur
  \piParL{a'}{E'}{Q}
}
\and
\inferrule*
{
  E \concur E'
  \\
  F \concur F'
}
{
  \piParLTau{E}{F}{y}
  \concur
  \piParLTau{E'}{F'}{z}
}
\and
\inferrule*
{
  E \concur E'
  \\
  F \concur F'
}
{
  \piParLTau{E}{F}{y}
  \concur
  \piParRTau{E'}{F'}{z}
}
\and
\inferrule*
{
  E \concur E'
  \\
  F \concur F'
}
{
  \piParLTau{E}{F}{y}
  \concur
  \piParLNu{E'}{F'}
}
\and
\inferrule*
{
  E \concur E'
  \\
  F \concur F'
}
{
  \piParLTau{E}{F}{y}
  \concur
  \piParRNu{E'}{F'}
}
\and
\inferrule*
{
  E \concur E'
  \\
  F \concur F'
}
{
  \piParLNu{E}{F}
  \concur
  \piParLNu{E'}{F'}
}
\and
\inferrule*
{
  E \concur E'
  \\
  F \concur F'
}
{
  \piParLNu{E}{F}
  \concur
  \piParRNu{E'}{F'}
}
\and
\inferrule*
{
  E \concur E'
}
{
  \piRestrictOutput{E}
  \concur
  \piRestrictOutput{E'}
}
\and
\inferrule*
{
  E \concur E'
}
{
  \piRestrictOutput{E}
  \concur
  \piRestrictA{a}{E'}
}
\and
\inferrule*
{
  E \concur E'
}
{
  \piRestrictA{a}{E}
  \concur
  \piRestrictA{a'}{E'}
}
\and
\inferrule*
{
  E \concur E'
}
{
  \piReplicate{E}
  \concur
  \piReplicate{E'}
}
\end{smathpar}
\crossrule
\caption{Concurrent coinitial transitions ($\piChoice{P}{\param}$, and
  some $\piParRTau{\param}{\param}{y}$ and $\piParRNu{\param}{\param}$
  variants omitted)}
\label{fig:concurrent}
\end{figure}

%% file: fig/residual.tex
\begin{figure}[H]
\noindent \shadebox{$\residual{E}{E'}$}
\begin{minipage}[t]{0.41\linewidth}
{\small
\begin{align*}
\residual{(\piParR{a}{P}{F})}
         {(\piParL{c}{E}{Q})}
&=
\piParR{a}{\target{E}}{F}
\\
\residual{(\piParR{a}{P}{F})}
         {(\piParL{b}{E}{Q})}
&=
\piParR{a}{\target{E}}{\ren{\push{}}{F}}
\\
\residual{(\piParL{a}{E}{Q})}
         {(\piParR{c}{P}{F})}
&=
\piParL{a}{E}{\target{F}}
\\
\residual{(\piParL{a}{E}{Q})}
         {(\piParR{b}{P}{F})}
&=
\piParL{a}{\ren{\push{}}{E}}{\target{F}}
\\
\residual{(\piParL{a}{E}{Q})}
         {(\piParLTau{E'}{F}{y})}
&=
\piParL{a}{\ren{(\pop{}{y})}{(\residual{E}{E'})}}{\target{F}}
\\
\residual{(\piParR{a}{P}{F})}
         {(\piParLTau{E}{F'}{y})}
&=
\piParR{a}{\ren{(\pop{}{y})}{\target{E}}}{\residual{F}{F'}}
\\
\residual{(\piParLTau{E}{F}{y})}
         {(\piParL{b}{E'}{Q})}
&=
\piParLTau{\residual{E}{E'}}{\ren{\push{}}{F}}{y}
\\
\residual{(\piParLTau{E}{F}{y})}
         {(\piParL{c}{E'}{Q})}
&=
\piParLTau{\residual{E}{E'}}{F}{y}
\\
\residual{(\piParLTau{E}{F}{y})}
         {(\piParR{b}{P}{F'})}
&=
\piParLTau{\ren{\push{}}{E}}{\residual{F}{F'}}{y}
\\
\residual{(\piParLTau{E}{F}{y})}
         {(\piParR{c}{P}{F'})}
&=
\piParLTau{E}{\residual{F}{F'}}{y}
\\
\residual{(\piParL{\piInput{x}}{E}{Q})}
         {(\piParLNu{E'}{F})}
&=
\piRestrictA{\piInput{x}}{(\piParL{\piInput{x+1}}{\residual{E}{E'}}{\target{F}})}
\\
\residual{(\piParL{\piBoundOutput{x}}{E}{Q})}
         {(\piParLNu{E'}{F})}
&=
\piRestrictOutput{(\piParL{\piOutput{x+1}{0}}{\residual{E}{E'}}{\target{F}})}
\\
\residual{(\piParL{c}{E}{Q})}
         {(\piParLNu{E'}{F})}
&=
\piRestrictA{c}{(\piParL{\ren{\push{}}{c}}{\residual{E}{E'}}{\target{F}})}
\\
\residual{(\piParR{\piInput{x}}{P}{F})}
         {(\piParLNu{E}{F'})}
&=
\piRestrictA{\piInput{x}}{(\piParR{\piInput{x+1}}{\target{E}}{\residual{F}{F'}})}
\\
\residual{(\piParR{\piBoundOutput{x}}{P}{F})}
         {(\piParLNu{E}{F'})}
&=
\piRestrictOutput{(\piParR{\piOutput{x+1}{0}}{\target{E}}{\residual{F}{F'}})}
\\
\residual{(\piParR{c}{P}{F})}
         {(\piParLNu{E}{F'})}
&=
\piRestrictA{c}{(\piParR{\ren{\push{}}{c}}{\target{E}}{\residual{F}{F'}})}
\\
\residual{(\piParLNu{E}{F})}
         {(\piParL{b}{E'}{Q})}
&=
\piParLNu{\residual{E}{E'}}{\ren{\push{}}{F}}
\\
\residual{(\piParLNu{E}{F})}
         {(\piParL{c}{E'}{Q})}
&=
\piParLNu{\residual{E}{E'}}{F}
\\
\residual{(\piParLNu{E}{F})}
         {(\piParR{\piInput{x}}{P}{F'})}
&=
\piParLNu{\ren{\push{}}{E}}{\residual{F}{F'}}
\\
\residual{(\piParLNu{E}{F})}
         {(\piParR{\piBoundOutput{x}}{P}{F'})}
&=
\piParLTau{\ren{\push{}}{E}}{\residual{F}{F'}}{0}
\\
\residual{(\piParLNu{E}{F})}
         {(\piParR{c}{P}{F'})}
&=
\piParLNu{E}{\residual{F}{F'}}
\\
\residual{(\piChoice{E}{Q})}
         {(\piChoice{E'}{Q})}
&=
\residual{E}{E'}
\\
\residual{(\piParR{\piInput{x}}{P}{F})}
         {(\piParR{b}{P}{F'})}
&=
\piParR{\piInput{x}}{\ren{\push{}}{P}}{\residual{F}{F'}}
\end{align*}
}
\end{minipage}%
\begin{minipage}[t]{0.5\linewidth}
{\small
\begin{align*}
\residual{(\piParR{b}{P}{F})}
         {(\piParR{\piInput{x}}{P}{F'})}
&=
\piParR{b}{\ren{\push{}}{P}}{\residual{F}{F'}}
\\
\residual{(\piParR{\piBoundOutput{x}}{P}{F})}
         {(\piParR{\piBoundOutput{u}}{P}{F'})}
&=
\piParR{\piOutput{x+1}{0}}{\ren{\push{}}{P}}{\residual{F}{F'}}
\\
\residual{(\piParR{c}{P}{F})}
         {(\piParR{b}{P}{F'})}
&=
\piParR{c}{\ren{\push{}}{P}}{\residual{F}{F'}}
\\
\residual{(\piParR{a}{P}{F})}
         {(\piParR{c}{P}{F'})}
&=
\piParR{a}{P}{\residual{F}{F'}}
\\
\residual{(\piParL{\piInput{x}}{E}{Q})}
         {(\piParL{b}{E'}{Q})}
&=
\piParL{\piInput{x}}{\residual{E}{E'}}{\ren{\push{}}{Q}}
\\
\residual{(\piParL{b}{E}{Q})}
         {(\piParL{\piInput{x}}{E'}{Q})}
&=
\piParL{b}{\residual{E}{E'}}{\ren{\push{}}{Q}}
\\
\residual{(\piParL{\piBoundOutput{x}}{E}{Q})}
         {(\piParL{\piBoundOutput{u}}{E'}{Q})}
&=
\piParL{\piOutput{x+1}{0}}{\residual{E}{E'}}{\ren{\push{}}{Q}}
\\
\residual{(\piParL{c}{E}{Q})}
         {(\piParL{b}{E'}{Q})}
&=
\piParL{c}{\residual{E}{E'}}{\ren{\push{}}{Q}}
\\
\residual{(\piParL{a}{E}{Q})}
         {(\piParL{c}{E'}{Q})}
&=
\piParL{a}{\residual{E}{E'}}{Q}
\\
\residual{(\piParLTau{E}{F}{y})}
         {(\piParLTau{E'}{F'}{z})}
&=
\piParLTau{\ren{(\pop{}{z})}{(\residual{E}{E'})}}{\residual{F}{F'}}{y}
\\
\residual{(\piParLTau{E}{F}{y})}
         {(\piParLNu{E'}{F'})}
&=
\piRestrictA{\tau}{(\piParLTau{\residual{E}{E'}}{\residual{F}{F'}}{y})}
\\
\residual{(\piParLNu{E}{F})}
         {(\piParLTau{E'}{F'}{z})}
&=
\piParLNu{\ren{(\pop{}{z})}{(\residual{E}{E'})}}{\residual{F}{F'}}
\\
\residual{(\piParLNu{E}{F})}
         {(\piParLNu{E'}{F'})}
&=
\piRestrictA{\tau}{(\piParLNu{\residual{E}{E'}}{\residual{F}{F'}})}
\\
\residual{(\piRestrictOutput{E})}
         {(\piRestrictOutput{E'})}
&=
\residual{E}{E'}
\\
\residual{(\piRestrictOutput{E})}
         {(\piRestrictA{b}{E'})}
&=
\piRestrictOutput{\;\ren{\swapR{}}{(\residual{E}{E'})}}
\\
\residual{(\piRestrictOutput{E})}
         {(\piRestrictA{c}{E'})}
&=
\piRestrictOutput{\;\residual{E}{E'}}
\\
\residual{(\piRestrictA{b}{E})}
         {(\piRestrictOutput{E'})}
&=
\residual{E}{E'}
\\
\residual{(\piRestrictA{c}{E})}
         {(\piRestrictOutput{E'})}
&=
\residual{E}{E'}
\\
\residual{(\piRestrictA{b}{E})}
         {(\piRestrictA{b}{E'})}
&=
\piRestrict{\residual{E}{E'}}
\\
\residual{(\piRestrictA{c}{E})}
         {(\piRestrictA{b}{E'})}
&=
\piRestrictA{c}{\;\ren{\swapR{}}{(\residual{E}{E'})}}
\\
\residual{(\piRestrictA{b}{E})}
         {(\piRestrictA{c}{E'})}
&=
\piRestrictA{b}{\;\residual{E}{E'}}
\\
\residual{(\piRestrictA{c}{E})}
         {(\piRestrictA{c}{E'})}
&=
\piRestrictA{c}{\;\residual{E}{E'}}
\\
\residual{(\piReplicate{E})}
         {(\piReplicate{E'})}
&=
\residual{E}{E'}
\end{align*}
}
\end{minipage} \\
\crossrule
\caption{Residual of $E$ after $E'$, omitting
  $\piParRTau{\param}{\param}{y}$ and $\piParRNu{\param}{\param}$ cases}
\label{fig:residual}
\end{figure}

%% file: fig/residual/illustrate/inputs-1.tex
\begin{center}
\scalebox{\smathparscale}{
\begin{smathpar}
\inferrule*[left={\ruleName{$\piParL{\piInput{x}}{\param}{\piAction{\piInput{z}}{Q}}$}}]
{
  \Gamma \vdash \piAction{\piInput{x}}{P}
  \transitionWithoutSmash{\piInput{x}}
  \suc{\Gamma} \vdash P
}
{
  \Gamma \vdash \piPar{\piAction{\piInput{x}}{P}}{\piAction{\piInput{z}}{Q}}
  \transitionWithoutSmash{\piInput{x}}
  \suc{\Gamma} \vdash \piPar{P}{\piAction{\piInput{\plus{z}{1}}}{\ren{(\suc{\push{}})}{Q}}}
}
\end{smathpar}
}
\end{center}

%% file: fig/residual/illustrate/inputs-2.tex
\begin{center}
\scalebox{\smathparscale}{
\begin{smathpar}
\inferrule*[left={\ruleName{$\piParR{\piInput{\suc{z}}}{P}{\param}$}}]
{
  \suc{\Gamma} \vdash \piAction{\piInput{\plus{z}{1}}}{\ren{(\suc{\pushR})}{Q}}
  \transitionWithoutSmash{\piInput{\suc{z}}}
  \plus{\Gamma}{2} \vdash \ren{(\suc{\pushR})}{Q}
}
{
  \suc{\Gamma} \vdash \piPar{P}{\piAction{\piInput{\plus{z}{1}}}{\ren{(\suc{\pushR})}{Q}}}
  \transitionWithoutSmash{\piInput{\suc{z}}}
  \plus{\Gamma}{2} \vdash \piPar{\ren{\pushR}{P}}{\ren{(\suc{\pushR})}{Q}}
}
\end{smathpar}
}
\end{center}

%% file: fig/residual/illustrate/inputs.tex
\begin{center}
\scalebox{0.8}{
\begin{tikzpicture}[node distance=1.5cm, auto]
  \node (PQ) [node distance=2cm] {
    $\Gamma \vdash \piPar{\piAction{\piInput{x}}{P}}{\piAction{\piInput{z}}{Q}}$
  };
  \node (PPushQ) [right of=PQ, above of=PQ] {
    $\plus{\Gamma}{1} \vdash \piPar{P}{\piAction{\piInput{\plus{z}{1}}}{\ren{(\suc{\push{}})}{Q}}}$
  };
  \node (PushPQ) [below of=PQ, right of=PQ] {
    $\plus{\Gamma}{1} \vdash \piPar{\piAction{\piInput{\plus{x}{1}}}{\ren{(\suc{\push{}})}{P}}}{Q}$
  };
  \node (PushPSucPushQ) [node distance=6.4cm, right of=PPushQ] {
    $\plus{\Gamma}{2} \vdash \piPar{\ren{\push{}}{P}}{\ren{(\suc{\push{}})}{Q}}$
  };
  \node (SwapPSwapQ) [below of=PushPSucPushQ] {
    $\plus{\Gamma}{2} \vdash \piPar{\ren{\swapR}{\ren{\pushR}{P}}}{\ren{\swapR}{\ren{(\suc{\pushR})}{Q}}}$
  };
  \node (SucPushPPushQ) [node distance=6.4cm, right of=PushPQ] {
    $\plus{\Gamma}{2} \vdash \piPar{\ren{(\suc{\push{}})}{P}}{\ren{\push{}}{Q}}$
  };
  \draw[->] (PQ) to node [yshift=-1ex] {$\piInput{x}$} (PPushQ);
  \draw[->] (PQ) to node [xshift=-1ex,yshift=2ex,swap] {$\piInput{z}$} (PushPQ);
  \draw[->] (PPushQ) to node {$\piInput{\suc{z}}$} (PushPSucPushQ);
  \draw[->] (PushPQ) to node [swap] {$\piInput{\suc{x}}$} (SucPushPPushQ);
  \draw[->] (PushPSucPushQ) to node {$\ren{\swapR}{}$} (SwapPSwapQ);
  \draw[-,double distance=1pt] (SwapPSwapQ) to node {$\piPar{\alpha}{\beta}$} (SucPushPPushQ);
\end{tikzpicture}
}
\end{center}

%% file: fig/congruence.tex
\begin{figure}[h]
\noindent \shadebox{$P \congEq R$}
\begin{smathpar}
\inferrule*[left={\ruleName{$\congRestrictSwap{P}$}}]
{
}
{
   \piRestrict{\piRestrict{(\ren{\swapR{}}{P}})}
   \congEq
   \piRestrict{\piRestrict{P}}
}
\and
\inferrule*[left={\ruleName{$\congRestrictSwapInv{P}$}}]
{
}
{
   \piRestrict{\piRestrict{P}}
   \congEq
   \piRestrict{\piRestrict{(\ren{\swapR{}}{P}})}
}
\and
\inferrule*[left={\ruleName{$\congTrans{\param}{\param}$}}]
{
   R \congEq S
   \\
   P \congEq R
}
{
   P \congEq S
}
\and
\inferrule*[left={\ruleName{$\piZero$}}]
{
}
{
   \piZero \congEq \piZero
}
\and
\inferrule*[left={\ruleName{$\piAction{\piInput{x}}{\param}$}}]
{
   P \congEq R
}
{
   \piAction{\piInput{x}}{P}
   \congEq
   \piAction{\piInput{x}}{R}
}
\and
\inferrule*[left={\ruleName{$\piAction{\piOutput{x}{y}}{\param}$}}]
{
   P \congEq R
}
{
   \piAction{\piOutput{x}{y}}{P}
   \congEq
   \piAction{\piOutput{x}{y}}{R}
}
\and
\inferrule*[left={\ruleName{$\piChoice{\param}{\param}$}}]
{
   P \congEq R
   \\
   Q \congEq S
}
{
   \piChoice{P}{Q}
   \congEq
   \piChoice{R}{S}
}
\and
\inferrule*[left={\ruleName{$\piPar{\param}{\param}$}}]
{
   P \congEq R
   \\
   Q \congEq S
}
{
   \piPar{P}{Q}
   \congEq
   \piPar{R}{S}
}
\and
\inferrule*[left={\ruleName{$\piRestrict{\param}$}}]
{
   P \congEq R
}
{
   \piRestrict{P}
   \congEq
   \piRestrict{R}
}
\and
\inferrule*[left={\ruleName{$\piReplicate{\param}$}}]
{
   P \congEq R
}
{
   \piReplicate{P}
   \congEq
   \piReplicate{R}
}
\end{smathpar}
\crossrule
\caption{Braiding congruence $\congEq$}
\label{fig:congruence}
\end{figure}

%% file: fig/renaming/congruence.tex
\scalebox{0.8}{
\begin{tikzpicture}[node distance=1.5cm, auto]
  \node (P) {$P$};
  \node (Q) [below of=P] {$\ren{\rho}{P}$};
  \node (PPrime) [node distance=3.4cm, right of=P] {$Q$};
  \node (QPrime) [below of=PPrime] {$\ren{\rho}{Q}$};
  \draw[->] (P) to node {$\phi$} (PPrime);
  \draw[dotted,->] (Q) to node [swap] {$\residual{\phi}{\rho}$} (QPrime);
  \draw[->] (P) to node [swap] {$\rho$} (Q);
  \draw[->] (PPrime) to node {$\rho$} (QPrime);
\end{tikzpicture}
}

%% file: fig/braiding.tex
\begin{minipage}[t]{0.4\linewidth}
{\small
\begin{align*}
\braid{\Gamma,0}
&=
\id_{\Gamma}: \Gamma \to \Gamma
\\
\braid{\Gamma,1}
&=
\id_{\suc{\Gamma}}: \suc{\Gamma} \to \suc{\Gamma}
\\
\braid{\Gamma,2}
&=
\swapR_{\Gamma}: \plus{\Gamma}{2} \to \plus{\Gamma}{2}
\end{align*}
}
\end{minipage}%
\begin{minipage}[t]{0.4\linewidth}
{\small
\begin{align*}
\\
P \cofinNew{\Gamma,\Delta} P'
& \iff
\ren{\braid{\Gamma,\Delta}}{P} \congEq P'
\end{align*}
}
\end{minipage}

%% file: fig/residual/illustrate/square.tex
\begin{nscenter}
\scalebox{0.8}{
\begin{tikzpicture}[node distance=1.5cm, auto]
  \node (P) [node distance=2cm] {
    $P$
  };
  \node (R) [right of=P, above of=P] {
    $R$
  };
  \node (RPrime) [below of=P, right of=P] {
    $R'$
  };
  \node (PPrime) [right of=R, below of=R] {
    $Q$
  };
  \draw[->] (P) to node [yshift=-1ex] {$E$} (R);
  \draw[->] (P) to node [yshift=1ex,swap] {$E'$} (RPrime);
  \draw[->] (R) to node {$\residual{E'}{E}$} (PPrime);
  \draw[->] (RPrime) to node [swap] {$\residual{E}{E'}$} (PPrime);
\end{tikzpicture}
}
\end{nscenter}

%% file: fig/residual/illustrate/pentagon.tex
\begin{nscenter}
\scalebox{0.8}{
\begin{tikzpicture}[node distance=1.5cm, auto]
  \node (P) [node distance=2cm] {
    $\Gamma \vdash P$
  };
  \node (R) [right of=P, above of=P] {
    $\Gamma' \vdash R$
  };
  \node (RPrime) [below of=P, right of=P] {
    $\Gamma'' \vdash R'$
  };
  \node (Q) [node distance=3.25cm, right of=R] {
    $\plus{\Gamma}{\Delta} \vdash Q$
  };
  \node (QPrime) [node distance=3.25cm, right of=RPrime] {
    $\plus{\Gamma}{\Delta} \vdash Q'$
  };
  \draw[->] (P) to node [yshift=-1ex] {$E$} (R);
  \draw[->] (P) to node [yshift=1ex,swap] {$E'$} (RPrime);
  \draw[->] (R) to node {$\residual{E'}{E}$} (Q);
  \draw[->] (RPrime) to node [swap] {$\residual{E}{E'}$} (QPrime);
  \draw[dotted, ->] (Q) to node {$\gamma$} (QPrime);
\end{tikzpicture}
}
\end{nscenter}

%% file: fig/concurrency-cube.tex
\begin{figure}[h]
{\small
\begin{nscenter}
\begin{tikzpicture}[
back line/.style={densely dotted},
cross line/.style={preaction={draw=white, -,
line width=6pt}}]
\matrix (m) [matrix of math nodes,
row sep=2em, column sep=1em,
text height=1.5ex,
text depth=0.25ex]{
& P & & R'' \\
R & & \mathbf{S_3} \\
& R' & & \mathbf{S_1} \\
\mathbf{S_2} & & \mathbf{P'} \\
};
\path[->,every node/.style={midway,sloped,above,font=\footnotesize}]
(m-1-2) edge node {$E'$} (m-1-4)
edge node {$E$} (m-2-1)
edge [back line] node [near end, xshift=1em, sloped=false] {$E''$} (m-3-2)
(m-1-4) edge node [sloped=false, xshift=1.25em] {$\residual{E''}{E}$} (m-3-4)
edge node [near end] {$\residual{E}{E'}$} (m-2-3)
(m-2-1) edge [cross line] node {$\residual{E'}{E}$} (m-2-3)
edge node [sloped=false, xshift=-1.25em] {$\residual{E''}{E}$} (m-4-1)
(m-3-2) edge [back line] (m-3-4)
edge [back line] (m-4-1)
(m-4-1) edge (m-4-3)
(m-3-4) edge (m-4-3)
(m-2-3) edge [cross line] node [near end, sloped=false, xshift=-1.25em, yshift=-1em]
   {$\residual{\mathbf{E''}}{\mathbf{\chi}}$} (m-4-3);
\end{tikzpicture}
\end{nscenter}
}
\end{figure}

%% file: sec/causal-equivalence.tex
\section{Causal equivalence}
\label{sec:causal-equivalence}

\subsection{Traces}

Define $\actions{\Gamma}$ to be the set of finite sequences of
composable actions starting at $\Gamma$. The empty sequence at $\Gamma$
is written $\asEmpty$; extension to the left is written
$\asCons{a}{\vec{a}}$. A \emph{trace} $t: P \transition{\vec{a}} R$ is a
finite sequence of composable transitions with initial state $\source{t}
= P$ and final state $\target{t} = R$. The empty trace at $P$ is written
$\sub{\trEmpty}{P}$; extension to the left of $t: R \transition{\vec{a}}
S$ by $E: P \transition{a} R$ is written $\trCons{E}{t}$.

\subsection{Residuals of traces and braidings}

To define the residual of a trace $t$ with respect to a braiding
$\gamma$, we first observe that a braiding congruence $\phi: P \congEq
P'$ commutes (on the nose) with a transition $E: P \transition{a} Q$,
inducing the corresponding notions of residual $\residual{\phi}{E}$ (the
image of the braiding congruence in the transition) and
$\residual{E}{\phi}$ (the image of the transition in the braiding
congruence).

\begin{theorem}
\label{thm:causal-equivalence:residual:braiding-congruence}
Suppose $E: P \transition{a} R$ and $\phi: P \congEq P'$. Then there
exists a process $R'$, transition $\residual{E}{\phi}: P' \transition{a}
R'$ and structural congruence $\residual{\phi}{E}: R \congEq R'$.

\input{fig/residual/congruence/cofinal}
\end{theorem}

\noindent \emph{Proof.} By the defining equations in
\figref{residual:congruence}.

Unlike residuals of the form $\residual{E}{E'}$, the cofinality of
$\residual{E}{\phi}$ and $\residual{\phi}{E}$ is by construction.
\appref{additional-proofs:causal-equivalence:residual:braiding-congruence}
illustrates cofinality for the cases where $\phi$ is of the form
$\congRestrictSwap{P}$.

\input{fig/residual/congruence}

To extend this notion of residuation from braiding congruences to
braidings requires a more general notion of braiding which permits the
renaming component of the braiding to be shifted under a binder. First
recall (from \defref{concurrent-transitions:cofin}) that any braiding
$\gamma: P \cofinNew{\Gamma, \Delta} P'$ is of the form $\phi \after
\braid{\Gamma,\Delta}$, where $\braid{\Gamma,\Delta}:
\plus{\Gamma}{\Delta} \to \plus{\Gamma}{\Delta}$ is the renaming $\id$
or $\swapR$, as determined by $\Delta \in \{0, 1, 2\}$, and $\phi$ is a
braiding congruence. We omit the $\Gamma,\Delta$ subscripts whenever
possible. The more general form of braiding allows the $\braid{}$ and
$\phi$ components to be translated by an arbitrary context $\Delta'$.

\begin{definition}[$\Delta$-shifted braiding]
For any context $\Delta$ define
\[
P \cofinVert{\Delta}{\Gamma,\Delta'} P'
\Longleftrightarrow
\ren{(\plus{\braid{\Gamma,\Delta'}}{\Delta})}{P} \congEq P'
\]
\end{definition}

Now we define the residual of a transition $E: \Gamma \vdash P
\transition{a} \plus{\Gamma}{\Delta} \vdash R$, where $\Delta \in \{0,
1\}$, and coinitial braiding $\gamma$ and show that the residual
$\residual{\gamma}{E}$ is $\gamma$ shifted by $\Delta$.

\begin{definition}[Residuals of transitions and braidings]
For any transition $E: P \transition{a} R$ and braiding $\gamma : P
\cofinVert{\Delta}{} P'$ with $\gamma = \phi \after \sigma$, define
$\residual{E}{\gamma}$ and $\residual{\gamma}{E}$ by the following
equations.

\input{fig/residual/braiding}

\noindent Cofinality is immediate by composing the square obtained by
applying \lemref{concurrent-transitions:renaming:transition} to $E$ and
$\sigma$ with the square obtained from
\thmref{causal-equivalence:residual:braiding-congruence} above to $\phi$
and $\residual{E}{\sigma}$. Closure of ($\Delta$-shifted) braidings
under residuation follows from the fact that $\residual{\sigma}{a} =
\plus{\sigma}{\Delta'}$ for some $\Delta' \in \{0, 1\}$.

\input{fig/residual/braiding/cofinal}

\noindent where both $\residual{E}{(\plus{\sigma}{\Delta})}$ and
$\residual{(\residual{E}{(\plus{\sigma}{\Delta})})}{\phi}$ have the action
$\ren{(\plus{\sigma}{\Delta})}{a}$.

Finally, we extend the definition to traces.

\begin{definition}[Residuals of action sequences and renamings]
\item Suppose $\rho: \Gamma \to \Delta$ and $\vec{a}: \actions{\Gamma}$.
  Define the residuals $\residual{\rho}{\vec{a}}$ and
  $\residual{\vec{a}}{\rho}$, writing the latter as
  $\ren{\rho}{\vec{a}}$.

\vspace{-16pt} 
\input{fig/residual/renaming/actions}
\vspace{4pt} 
\end{definition}

\begin{lemma}[Residuals of traces and braidings]
\item
Suppose $t: P \transition{\vec{a}} R$ and $\gamma = \phi \after \sigma:
P \cofinVert{\Delta}{} P'$. Then there exists a process $R'$, trace $P'
\transition{\ren{\sigma}{\vec{a}}} R'$ and braiding
$\residual{\gamma}{t}: R \cofinNew{} R'$.
\input{fig/residual/braiding/sequence/cofinal}
\end{lemma}

\noindent \emph{Proof.} By the following defining equations.

\input{fig/residual/braiding/sequence}

\subsection{Causal equivalence}

We now define \emph{causal equivalence}, the congruence over traces
induced by the notion of transition residual from
\secref{concurrent-transitions:residuals}. A causal equivalence $\alpha:
t \permEq u$ witnesses the reordering of one trace $t$ into a coinitial
trace $u$ by the permutation of concurrent transitions. Meta-variables
$\alpha$, $\beta$ range over causal equivalences.

\begin{definition}
Inductively define the relation $\permEq$ given by the rules in
\figref{causal-equivalence}, where syntactically $\permEq$ has lower
priority than $\trCons{\param}{\param}$. If $\alpha : t \permEq u$ then
$\source{\alpha}$ and $\target{\alpha}$ denote $t$ and $u$ respectively.

\input{fig/causal-equivalence}

\end{definition}
The $\sub{\trEmpty}{P}$ and $\trCons{E}{\alpha}$ rules are the
congruence cases. The $\permEqTrans{\alpha}{\beta}$ rule closes under
transitivity, which is a form of vertical composition. The transposition
rule $\permEqSwap{E}{E'}{\alpha}$ extends an existing causal equivalence
$\alpha: t \permEq u$ with the two possible interleavings of concurrent
steps $E \concur E'$. What is interesting about this rule is that the
trace $u$ must be transported through the braiding $\rewire{E}{E'}$
witnessing the cofinality of $E$ and $E'$, in order to obtain a trace
$\residual{u}{\rewire{E}{E'}}$ composable with $\residual{E'}{E}$. The
following diagram illustrates.

\input{fig/transpose-rule}

As the diagram suggests, the transposition rule causes braidings to
compose vertically. Here, $\rewireVec{\alpha}$ is a composite braiding
relating $S$ to $S'$, which is extended by the braiding
$\residual{\rewire{E}{E'}}{u}$ to relate $S$ to $S''$. We leave
formalising this aspect of causal equivalence to future work.

\begin{theorem}
$\permEq$ is an equivalence relation.
\end{theorem}

\noindent \emph{Proof.} Reflexivity is a trivial induction, using the
$\sub{\trEmpty}{P}$ and $\trCons{E}{\alpha}$ rules. Transitivity is
immediate from the $\permEqTrans{\alpha}{\beta}$ rule. Symmetry is
trivial in the $\sub{\trEmpty}{P}$, $\trCons{E}{\alpha}$ and
$\permEqTrans{\alpha}{\beta}$ cases. The $\permEqSwap{E}{E'}{\alpha}$
case requires the symmetry of $\concur$ and that
$\residual{(\residual{u}{\rewireVec{\alpha}})}{\rewireVec{\alpha}^{-1}}
= u$, where $u = \target{\alpha}$.

%% file: fig/residual/congruence/cofinal.tex
\begin{nscenter}
\scalebox{0.8}{
\begin{tikzpicture}[node distance=1.5cm, auto]
  \node (P) {$P$};
  \node (PPrime) [below of=P] {$P'$};
  \node (R) [node distance=3cm, right of=P] {$R$};
  \node (RPrime) [below of=R] {$R'$};
  \draw[->] (P) to node {$E$} (R);
  \draw[dotted,->] (PPrime) to node [swap] {$\residual{E}{\phi}$} (RPrime);
  \draw[->] (P) to node [swap] {$\phi$} (PPrime);
  \draw[dotted,->] (R) to node {$\residual{\phi}{E}$} (RPrime);
\end{tikzpicture}
}
\end{nscenter}

%% file: fig/residual/congruence.tex
\begin{figure}[h]
\begin{nscenter}
\begin{minipage}[t]{0.49\linewidth}
\noindent \shadebox{$\residual{E}{\phi}$}
{\small
\begin{align*}
\smash{\residual{\congRestrictSwap{\source{E}}}{(\piRestrictOutput{\piRestrictA{\piOutput{\suc{x}}{0}}{E}})}}
&=
\smash{\piRestrictA{\piBoundOutput{x}}{\piRestrictOutput{(\ren{\swapR}{E})}}}
\\
\smash{\residual{\congRestrictSwap{\source{E}}}{(\piRestrictA{\piBoundOutput{x}}{\piRestrictOutput{E}})}}
&=
\smash{\piRestrictOutput{\piRestrictA{\piOutput{x+1}{0}}{(\ren{\swapR}{E})}}}
\\
\smash{\residual{\congRestrictSwap{\source{E}}}{(\piRestrictA{c}{\piRestrictA{c'}{E}})}}
&=
\smash{\piRestrictA{c}{\piRestrictA{c'}{(\ren{\swapR}{E}})}}
\\
\smash{\residual{\congRestrictSwap{\source{E}}}{(\piRestrictA{b}{\piRestrictA{b'}{E}})}}
&=
\smash{\piRestrictA{b}{\piRestrictA{b'}{(\ren{\swapR}{E}})}}
\\
\residual{(\piAction{\piInput{x}}{P})}{(\piAction{\piInput{x}}{\phi})}
&=
\piAction{\piInput{x}}{\target{\phi}}
\\
\residual{(\piAction{\piOutput{x}{y}}{P})}{(\piAction{\piOutput{x}{y}}{\phi})}
&=
\piAction{\piOutput{x}{y}}{\target{\phi}}
\\
\residual{(\piChoice{E}{Q})}{(\piChoice{\phi}{\psi})}
&=
\piChoice{\residual{E}{\phi}}{\target \psi}
\\
\residual{(\piParL{b}{E}{Q})}{(\piChoice{\phi}{\psi})}
&=
\piParL{b}{\residual{E}{\phi}}{\target \psi}
\\
\residual{(\piParL{c}{E}{Q})}{(\piChoice{\phi}{\psi})}
&=
\piParL{c}{\residual{E}{\phi}}{\target \psi}
\\
\residual{(\piParR{b}{P}{F})}{(\piChoice{\phi}{\psi})}
&=
\piParR{b}{\target \phi}{\residual{F}{\psi}}
\\
\residual{(\piParR{c}{P}{F})}{(\piChoice{\phi}{\psi})}
&=
\piParR{c}{\target \phi}{\residual{F}{\psi}}
\\
\residual{(\piParLTau{E}{F}{y})}{(\piPar{\phi}{\psi})}
&=
\piParLTau{\residual{E}{\phi}}{\residual{F}{\psi}}{y}
\\
\residual{(\piParLNu{E}{F})}{(\piPar{\phi}{\psi})}
&=
\piParLNu{\residual{E}{\phi}}{\residual{F}{\psi}}
\\
\residual{(\piRestrictOutput{E})}{(\piRestrict{\phi})}
&=
\piRestrictOutput{\;\residual{E}{\phi}}
\\
\residual{(\piRestrictA{b}{E})}{(\piRestrict{\phi})}
&=
\piRestrictA{b}{\residual{E}{\phi}}
\\
\residual{(\piRestrictA{c}{E})}{(\piRestrict{\phi})}
&=
\piRestrictA{c}{\residual{E}{\phi}}
\\
\residual{(\piReplicate{E})}{(\piReplicate{\phi})}
&=
\piReplicate{\residual{E}{(\piPar{\phi}{\piReplicate{\phi}})}}
\\
\residual{E}{(\congTrans{\phi'}{\phi})}
&=
\residual{(\residual{E}{\phi})}{\phi'}
\end{align*}}
\end{minipage}%
\begin{minipage}[t]{0.49\linewidth}
\noindent \shadebox{$\residual{\phi}{E}$}
{\small
\begin{align*}
\smash{\residual{\congRestrictSwap{\source{E}}}{(\piRestrictOutput{\piRestrictA{\piOutput{\suc{x}}{0}}{E}})}}
&=
\piRestrict{\;\congRefl{\target{E}}}
\\
\smash{\residual{\congRestrictSwap{\source{E}}}{(\piRestrictA{\piBoundOutput{x}}{\piRestrictOutput{E}})}}
&=
\piRestrict{\;\congRefl{\ren{\swapR}{\target{E}}}}
\\
\smash{\residual{\congRestrictSwap{\source{E}}}{(\piRestrictA{c}{\piRestrictA{c'}{E}})}}
&=
\smash{\congRestrictSwapInv{\target{E}}}
\\
\smash{\residual{\congRestrictSwap{\source{E}}}{(\piRestrictA{b}{\piRestrictA{b'}{E}})}}
&=
\smash{\congRestrictSwapInv{\ren{\swapR}{\ren{(\suc{\swapR})}{\ren{\swapR}{\target{E}}}}}}
\\
\smash{\residual{(\piAction{\piInput{x}}{\phi})}{(\piAction{\piInput{x}}{P})}}
&=
\phi
\\
\residual{(\piAction{\piOutput{x}{y}}{\phi})}{(\piAction{\piOutput{x}{y}}{P})}
&=
\phi
\\
\residual{(\piChoice{\phi}{\psi})}{(\piChoice{E}{Q})}
&=
\residual{\phi}{E}
\\
\residual{(\piChoice{\phi}{\psi})}{(\piParL{b}{E}{Q})}
&=
\piPar{\residual{\phi}{E}}{\ren{\push{}}{\psi}}
\\
\residual{(\piChoice{\phi}{\psi})}{(\piParL{c}{E}{Q})}
&=
\piPar{\residual{\phi}{E}}{\psi}
\\
\residual{(\piChoice{\phi}{\psi})}{(\piParR{b}{P}{F})}
&=
\piPar{\ren{\push{}}{\phi}}{\residual{\psi}{F}}
\\
\residual{(\piChoice{\phi}{\psi})}{(\piParR{c}{P}{F})}
&=
\piPar{\phi}{\residual{\psi}{F}}
\\
\residual{(\piPar{\phi}{\psi})}{(\piParLTau{E}{F}{y})}
&=
\ren{(\pop{}{y})}{\residual{\phi}{E}}
\\
\residual{(\piPar{\phi}{\psi})}{(\piParLNu{E}{F})}
&=
\piRestrict{(\piPar{\residual{\phi}{E}}{\residual{\psi}{F}})}
\\
\residual{(\piRestrict{\phi})}{(\piRestrictOutput{E})}
&=
\residual{\phi}{E}
\\
\residual{(\piRestrict{\phi})}{(\piRestrictA{b}{E})}
&=
\piRestrict{\;\ren{\swapR{}}{\residual{\phi}{E}}}
\\
\residual{(\piRestrict{\phi})}{(\piRestrictA{c}{E})}
&=
\piRestrict{\;\residual{\phi}{E}}
\\
\residual{(\piReplicate{\phi})}{(\piReplicate{E})}
&=
\residual{(\piPar{\phi}{\piReplicate{\phi}})}{E}
\\
\residual{(\congTrans{\phi'}{\phi})}{E}
&=
\congTrans{(\residual{\phi'}{(\residual{E}{\phi})})}{\residual{\phi}{E}}
\end{align*}}
\end{minipage}
\end{nscenter}
\crossrule
\caption{Residual of transition $E$ and coinitial braiding congruence
  $\phi$}
\label{fig:residual:congruence}
\end{figure}

%% file: fig/residual/braiding.tex
\begin{minipage}[t]{0.4\linewidth}
\begin{nscenter}
{\small
\begin{align*}
\residual{E}{(\phi \after \sigma)}
&=
\residual{(\residual{E}{\sigma})}{\phi}
\end{align*}
}
\end{nscenter}
\end{minipage}%
\begin{minipage}[t]{0.4\linewidth}
\begin{nscenter}
{\small
\begin{align*}
\residual{(\phi \after \sigma)}{E}
&=
(\residual{\phi}{(\residual{E}{\sigma})}) \after \residual{\sigma}{a}
\end{align*}
}
\end{nscenter}
\end{minipage}
\end{definition}

%% file: fig/residual/braiding/cofinal.tex
\begin{center}
\scalebox{0.8}{
\begin{tikzpicture}[node distance=1.5cm, auto]
  \node (A) {$P$};
  \node (B) [node distance=3.75cm, right of=A] {$R$};
  \node (P) [below of=A] {$\ren{(\plus{\sigma}{\Delta})}{P}$};
  \node (PPrime) [below of=P] {$P'$};
  \node (R) [node distance=3.75cm, right of=P] {$\ren{(\plus{\sigma}{\Delta'})}{R}$};
  \node (RPrime) [below of=R] {$R'$};
  \draw[->] (A) to node {$E$} (B);
  \draw[->] (A) to node [swap] {$\plus{\sigma}{\Delta}$} (P);
  \draw[->] (B) to node {$\plus{\sigma}{\Delta'}$} (R);
  \draw[->] (P) to node {$\residual{E}{(\plus{\sigma}{\Delta})}$} (R);
  \draw[->] (PPrime) to node [swap] {$\residual{(\residual{E}{(\plus{\sigma}{\Delta})})}{\phi}$} (RPrime);
  \draw[->] (P) to node [swap] {$\phi$} (PPrime);
  \draw[->] (R) to node {$\residual{\phi}{(\residual{E}{(\plus{\sigma}{\Delta})})}$} (RPrime);
\end{tikzpicture}
}
\end{center}

%% file: fig/residual/renaming/actions.tex
\begin{nscenter}
\begin{minipage}[t]{0.29\linewidth}
\begin{align*}
\residual{\rho}{\sub{\asEmpty}{\Gamma}}
&=
\rho
\\
\ren{\rho}{\sub{\asEmpty}{\Gamma}}
&=
\sub{\asEmpty}{\Delta}
\end{align*}
\end{minipage}%
\begin{minipage}[t]{0.29\linewidth}
\begin{align*}
\residual{\rho}{(\asCons{a}{\vec{a}})}
&=
\residual{(\residual{\rho}{a})}{\vec{a}}
\\
\ren{\rho}{(\asCons{a}{\vec{a}})}
&=
\asCons{(\ren{\rho}{a})}{\ren{(\residual{\rho}{a})}{\vec{a}}}
\end{align*}
\end{minipage}
\end{nscenter}

%% file: fig/residual/braiding/sequence/cofinal.tex
\begin{nscenter}
\scalebox{0.8}{
\begin{tikzpicture}[node distance=1.5cm, auto]
  \node (P) {$P$};
  \node (PPrime) [below of=P] {$P'$};
  \node (R) [node distance=3cm, right of=P] {$R$};
  \node (RPrime) [below of=R] {$R'$};
  \draw[->] (P) to node {$t$} (R);
  \draw[dotted,->] (PPrime) to node [swap] {$\residual{t}{\gamma}$} (RPrime);
  \draw[->] (P) to node [swap] {$\gamma$} (PPrime);
  \draw[dotted,->] (R) to node {$\residual{\gamma}{t}$} (RPrime);
\end{tikzpicture}
}
\end{nscenter}

%% file: fig/residual/braiding/sequence.tex
\begin{nscenter}
\begin{minipage}[t]{0.35\linewidth}
\begin{nscenter}
\scalebox{0.8}{
\begin{tikzpicture}[node distance=1.5cm, auto]
  \node (P) {$P$};
  \node (PPrime) [below of=P] {$P'$};
  \node (R) [node distance=2cm, right of=P] {$P$};
  \node (RPrime) [below of=R] {$P'$};
  \draw[-,double distance=1pt] (P) to node {$\trEmpty$} (R);
  \draw[-,double distance=1pt] (PPrime) to node [swap] {$\residual{\trEmpty}{\gamma}$} (RPrime);
  \draw[->] (P) to node [swap] {$\gamma$} (PPrime);
  \draw[->] (R) to node {$\residual{\gamma}{\trEmpty}$} (RPrime);
\end{tikzpicture}
}
\end{nscenter}
\end{minipage}%
\begin{minipage}[t]{0.35\linewidth}
\begin{nscenter}
\scalebox{0.8}{
\begin{tikzpicture}[node distance=1.5cm, auto]
  \node (P) {$P$};
  \node (PPrime) [below of=P] {$P'$};
  \node (R) [node distance=2.5cm, right of=P] {$R$};
  \node (RPrime) [below of=R] {$R'$};
  \node (S) [node distance=2.5cm, right of=R] {$S$};
  \node (SPrime) [below of=S] {$S'$};
  \draw[->] (P) to node {$E$} (R);
  \draw[->] (R) to node {$t$} (S);
  \draw[->] (PPrime) to node [swap] {$\residual{E}{\gamma}$} (RPrime);
  \draw[->] (RPrime) to node [swap] {$\residual{t}{(\residual{\gamma}{E})}$} (SPrime);
  \draw[->] (P) to node [swap] {$\gamma$} (PPrime);
  \draw[->] (R) to node {$\residual{\gamma}{E}$} (RPrime);
  \draw[->] (S) to node {$\residual{(\residual{\gamma}{E})}{t}$} (SPrime);
\end{tikzpicture}
}
\end{nscenter}
\end{minipage}
\begin{minipage}[t]{0.35\linewidth}
{\small
\begin{align*}
\residual{\sub{\trEmpty}{P}}{\gamma}
&=
\sub{\trEmpty}{P'}
\\
\residual{\gamma}{\sub{\trEmpty}{P}}
&=
\gamma
\end{align*}}
\end{minipage}%
\begin{minipage}[t]{0.35\linewidth}
{\small
\begin{align*}
\residual{(\trCons{E}{t})}{\gamma}
&=
\trCons{(\residual{E}{\gamma})}{\residual{t}{(\residual{\gamma}{E})}}
\\
\residual{\gamma}{(\trCons{E}{t})}
&=
\residual{(\residual{\gamma}{E})}{t}
\end{align*}}
\end{minipage}
\end{nscenter}

%% file: fig/causal-equivalence.tex
\begin{figure}[h]
\noindent \shadebox{$t \permEq u$}
\begin{smathpar}
\inferrule*[left={\ruleName{$\sub{\trEmpty}{P}$}}]
{
}
{
  \sub{\trEmpty}{P} \permEq \sub{\trEmpty}{P}
}
\and
\inferrule*[left={\ruleName{$\trCons{\param}{\param}$}},right={\textnormal{$\source{t} = R$}}]
{
  E: P \transition{a} R
  \\
  t \permEq u
}
{
  \trCons{E}{t} \permEq \trCons{E}{u}
}
\and
\inferrule*[left={\ruleName{$\permEqTrans{\param}{\param}$}}]
{
  t' \permEq u
  \\
  t \permEq t'
}
{
  t \permEq u
}
\and
\inferrule*[left={\ruleName{$\permEqSwap{\param}{\param}{\param}$}},right={$E \concur E'$}]
{
  E: P \transition{a} R
  \\
  E': P \transition{a'} R'
  \\
  t \permEq u
}
{
  \trCons{E}{\trCons{\residual{E'}{E}}{t}}
  \permEq
  \trCons{E'}{\trCons{\residual{E}{E'}}{\residual{u}{\rewire{E}{E'}}}}
}
\end{smathpar}
\crossrule
\caption{Causal equivalence}
\label{fig:causal-equivalence}
\end{figure}

%% file: fig/transpose-rule.tex
\begin{center}
\scalebox{0.8}{
\begin{tikzpicture}[node distance=1.5cm, auto]
  \node (P) {$P$};
  \node (R) [node distance=1.5cm, above of=P, right of=P] {$R$};
  \node (R') [node distance=1.5cm, below of=P, right of=P] {$R'$};
  \node (Q) [node distance=2.5cm, right of=R] {$Q$};
  \node (Q') [node distance=2.5cm, right of=R'] {$Q'$};
  \node (S) [node distance=2.5cm, right of=Q] {$S$};
  \node (S') [below of=S] {$S'$};
  \node (S'') [below of=S'] {$S''$};
  \draw[->] (P) to node {$E$} (R);
  \draw[->] (P) to node [swap] {$E'$} (R');
  \draw[->] (R) to node {$\residual{E'}{E}$} (Q);
  \draw[->] (R') to node [swap] {$\residual{E}{E'}$} (Q');
  \draw[->] (Q) to node {$t$} (S);
  \draw[->] (Q) to node [swap] {$u$} (S');
  \draw[->] (Q') to node [swap] {$\residual{u}{\rewire{E}{E'}}$} (S'');
  \draw[->] (Q) to node [swap] {$\rewire{E}{E'}$} (Q');
  \draw[->] (S) to node {$\rewireVec{\alpha}$} (S');
  \draw[->] (S') to node {$\residual{\rewire{E}{E'}}{u}$} (S'');
\end{tikzpicture}}
\end{center}

%% file: sec/related-work.tex
\section{Related work}
\label{sec:related-work}

Hirschkoff's $\mu s$ calculus \cite{hirschkoff99} has a similar
treatment of de Bruijn indices. Its renaming operators $\langle x
\rangle$, $\phi$ and $\psi$ are effectively our $\pop{}{x}$, $\push{}$
and $\swapR{}$ renamings, but fused with the $\ren{\param}{}$ operator
which applies a renaming to a process. Hirschkoff's operators are also
syntactic forms in the $\mu s$ calculus, rather than meta-operations,
and therefore the operational semantics also includes rules for reducing
occurrences of the renaming operators that arise during a process
reduction step.


Formalisations of the \piCalculus have been undertaken in several
theorem provers used for mechanised metatheory.
Due to space limits, we limit attention to closely-related
formalisation techniques based on constructive logics.


\newcommand{\typename}[1]{\textsf{#1}}

\Paragraph{Coq}
Hirschkoff~\cite{hirschkoff97} formalised the \piCalculus in Coq using
de Bruijn indices, and verified properties such as congruence and
structural equivalence laws of bisimulation.
Despeyroux~\cite{despeyroux00} formalised the \piCalculus in Coq using
weak higher-order abstract syntax, assuming a decidable type of names,
and using two separate transitions, for ordinary, input and output
transitions respectively; for input and output transitions the
right-hand side is a function of type $\typename{name} \to
\typename{proc}$. This formalisation included a simple type system and
proof of type soundness. Honsell, Miculan and Scagnetto~\cite{honsell01}
formalised the \piCalculus in Coq, also using weak higher-order abstract
syntax. The type of names \typename{name} is a type parameter assumed to
admit decidable equality and freshness ($\typename{notin}$) relations.
Transitions are encoded using two inductive definitions, for free and
bound actions, which differ in the type of the third argument
($\typename{proc}$ vs.~$\typename{name} \to \typename{proc}$). Numerous
results from Milner, Parrow and Walker~\cite{milner92} are verified,
using the \emph{theory of contexts} (whose axioms are assumed in their
formalisation, but have been validated semantically).

\Paragraph{CLF} Cervesato, Pfenning, Walker and Watkins~\cite{cervesato02}
formalise synchronous and asynchronous versions of \piCalculus in the
Concurrent Logical Framework (CLF).
CLF employs
higher-order abstract syntax, linearity and a monadic encapsulation of
certain linear constructs that can identify objects such as traces up to
causal equivalence. Thus, CLF's \piCalculus encodings naturally induce
equivalences on traces. However, a nontrivial effort appears necessary
to compare CLF's notion of trace equivalence with others (including
ours) due to the distinctive approach taken in CLF.

\Paragraph{Agda} Orchard and Yoshida~\cite{orchard15} present a
translation from a functional language with effects to a \piCalculus
with session types and verify some type-preservation properties of the
translation in Agda.


%% file: sec/conclusion.tex
\section{Conclusions and future work}
\label{sec:conclusion}

To the best of our knowledge, we are the first to report on a
formalisation of the operational behavior of the \piCalculus in Agda.
Compared to prior formalisations, ours is distinctive in two ways.

First, our formalisation employs an indexed family of types for process
terms and uses the indices instead of binding to deal with scope
extrusion. Formalisations of lambda-calculi often employ this technique,
but to our knowledge only Orchard and Yoshida report a similar approach
for a \piCalculus formalisation. This choice helps tame the complexity
of de Bruijn indices, because many invariants are automatically checked
by the type system rather than requiring additional explicit reasoning.

Second, our work appears to be the first to align the notion of ``proved
transitions'' from Boudol and Castellani's work on CCS with ``transition
proofs'' in the \piCalculus. This hinges on the capability to manipulate
and perform induction or recursion over derivations, and means we can
leverage dependent typing so that residuation is defined only for
concurrent transitions, rather than on all pairs of transitions. It is
worth noting that while CLF's approach to encoding \piCalculus
automatically yields an equivalence on traces, it is unclear (at least
to us) whether this equivalence is the same as the one we propose, or
whether such traces can be manipulated explicitly as proof objects if
desired.

In future work we may explore trace structures explicitly quotiented by
causal equivalence, such as dependence graphs \cite{mazurkiewicz87} or
event structures \cite{boudol89}. We are also interested in extending
braiding congruence to the full \piCalculus structural congruence, and
in understanding whether and how ideas from homotopy type
theory~\cite{HoTTbook}, such as quotients or higher inductive types,
could be applied to ease reasoning about or correct programming with
\piCalculus terms (modulo structural congruence) or traces (modulo
causal equivalence).

%% file: acknowledgements.tex
\paragraph*{Acknowledgements}

The authors were supported by the Air Force Office of Scientific
Research, Air Force Material Command, USAF, under grant number
FA8655-13-1-3006. The first author was also supported by UK EPSRC project
\emph{From Data Types to Session Types: A Basis for Concurrency and
Distribution} (EP/K034413/1).

%% file: sec/appendix.tex
\begin{appendix}

\input{sec/appendix/module-structure}

\input{sec/appendix/renaming-lemmas}
\input{sec/appendix/additional-proofs}

\end{appendix}

%% file: sec/appendix/module-structure.tex
\section{Agda module structure}
\label{app:module-structure}

\figref{modules} summarises the module structure of the Agda
formalisation.

\input{fig/modules.tex}

%% file: fig/modules.tex
\begin{figure}[H]
\begin{nscenter}
{\small
\noindent \begin{tabular}{lL{10cm}}
\text{\emph{Utilities}}
\\
\tt{Common}
& Useful definitions not found in the Agda standard library
\\
\tt{SharedModules}
& Common imports from standard library
\\
\\
\text{\emph{Core modules}}
\\
\tt{Action}
& Actions $a$
\\
\tt{Action.Concur}
& Concurrent actions $a \concur a'$; residuals $\residual{a}{a'}$
\\
\tt{Action.Concur.Action}
& Residual of $a \concur a'$ after $a''$
\\
\tt{Action.Seq}
& Action sequences $\vec{a}$
\\
\tt{Name}
& Contexts $\Gamma$; names $x$
\\
\tt{Proc}
& Processes $P$
\\
\tt{Ren}
& Renamings $\rho: \Gamma \to \Gamma'$
\\
\tt{StructuralCong.Proc}
& Braiding congruence relation $\phi: P \congEq P'$
\\
\tt{StructuralCong.Transition}
& Residuals $\residual{E}{\phi}$ and $\residual{\phi}{E}$
\\
\tt{Transition}
& Transitions $E: P \transition{a} R$
\\
\tt{Transition.Concur}
& Concurrent transitions $\chi: E \concur E'$; residuals $\residual{E}{E'}$
\\
\tt{Transition.Concur.Cofinal}
& Cofinality braidings $\gamma$
\\
\tt{Transition.Concur.Cofinal.Transition}
& Residuals $\residual{E}{\gamma}$ and $\residual{\gamma}{E}$
\\
\tt{Transition.Concur.Transition}
& Residual $\residual{\chi}{E}$
\\
\tt{Transition.Seq}
& Transition sequences
\\
\tt{Transition.Seq.Cofinal}
& Residuals $\residual{t}{\gamma}$ and $\residual{\gamma}{t}$; permutation equivalence $\alpha: t \permEq u$
\\
\\
\text{\emph{Typical sub-modules}}
\\
\tt{.Properties}
& Additional properties relating to $X$
\\
\tt{.Ren}
& Renaming lifted to $X$
\\ 
\\
\end{tabular}
}
\end{nscenter}
\crossrule
\caption{Module overview}
\label{fig:modules}
\end{figure}

%% file: sec/appendix/renaming-lemmas.tex
\section{Renaming lemmas}
\label{app:renaming-lemmas}

Each lemma asserts the commutativity of the diagram on the left; when a
string diagram is also provided, it should be interpreted as an informal
proof sketch.

\paragraph{\lemref{pop-after-push}.}
\begin{center}
\begin{minipage}{0.22\linewidth}
\input{fig/lemma/pop-after-push}
\end{minipage}
\begin{minipage}{0.67\linewidth}
\input{fig/lemma/illustrate/pop-after-push}
\end{minipage}
\end{center}

\paragraph{\lemref{pop-zero-after-suc-push}.}
\begin{center}
\begin{minipage}{0.25\linewidth}
\input{fig/lemma/pop-zero-after-suc-push}
\end{minipage}
\begin{minipage}{0.67\linewidth}
\input{fig/lemma/illustrate/pop-zero-after-suc-push}
\end{minipage}
\end{center}

\paragraph{\lemref{swap-after-suc-swap-after-swap}.}
\begin{center}
\begin{minipage}{0.42\linewidth}
\input{fig/lemma/swap-after-suc-swap-after-swap}
\end{minipage}
\begin{minipage}{0.52\linewidth}
\input{fig/lemma/illustrate/swap-after-suc-swap-after-swap}
\end{minipage}
\end{center}

\paragraph{\lemref{pop-swap}.}
\begin{center}
\begin{minipage}{0.38\linewidth}
\input{fig/lemma/pop-swap}
\end{minipage}
\begin{minipage}{0.52\linewidth}
\input{fig/lemma/illustrate/pop-swap}
\end{minipage}
\end{center}

\paragraph{\lemref{swap-push}.}
\begin{center}
\begin{minipage}{0.22\linewidth}
\input{fig/lemma/swap-push}
\end{minipage}
\begin{minipage}{0.68\linewidth}
\input{fig/lemma/illustrate/swap-push-1}
\vspace{-5pt}
\input{fig/lemma/illustrate/swap-push-2}
\end{minipage}
\end{center}

\paragraph{\lemrefThree{push-comm}{pop-comm}{swap-suc-suc}.}
\begin{center}
\begin{minipage}{0.5\linewidth}
\begin{nscenter}
\input{fig/lemma/push-comm-pop-comm}
\end{nscenter}
\end{minipage}
\begin{minipage}{0.3\linewidth}
\begin{nscenter}
\input{fig/lemma/swap-suc-suc}
\end{nscenter}
\end{minipage}
\end{center}

%% file: fig/lemma/pop-after-push.tex
\scalebox{0.8}{
\begin{tikzpicture}[node distance=1.5cm, auto, baseline={([yshift=-.5ex]current bounding box.center)}]
  \node (GammaA) {$\Gamma$};
  \node (GammaB) [node distance=3cm, right of=GammaA] {$\suc{\Gamma}$};
  \node (GammaC) [below of=GammaB] {$\Gamma$};
  \draw[->] (GammaA) to node {$\push{}$} (GammaB);
  \draw[-,double distance=1pt] (GammaA) to node [xshift=0.8em, swap] {} (GammaC);
  \draw[->, shift left=0.2em] (GammaB) to node {$\pop{}{x}$} (GammaC);
\end{tikzpicture}
}

%% file: fig/lemma/illustrate/pop-after-push.tex
\begin{center}
\scalebox{0.8}{
\begin{tikzpicture}[node distance=0.5cm, auto]
  \node [anchor=west] at (0,0) (LabelA) {$\Gamma$};
  \node [anchor=west, below of=LabelA] (0A) {0};
  \node [anchor=west, below of=0A] (1A) {1};
  \node [anchor=west, below of=1A] (2A) {$\vdots$};

  \node [anchor=east, right of=LabelA, node distance=2.5cm] (LabelB) {$\suc{\Gamma}$};
  \node [anchor=east, below of=LabelB] (0B) {0};
  \node [anchor=east, below of=0B] (1B) {1};
  \node [anchor=east, below of=1B] (2B) {2};
  \node [anchor=east, below of=2B] (3B) {$\vdots$};

  \node [anchor=east, right of=LabelB, node distance=2.5cm] (LabelC) {$\Gamma$};
  \node [anchor=east, below of=LabelC] (0C) {0};
  \node [anchor=east, below of=0C] (1C) {1};
  \node [anchor=east, below of=1C] (2C) {2};
  \node [anchor=east, below of=2C] (3C) {$\vdots$};
  \node [anchor=east, below of=3C] (4C) {$x$};

  \draw (LabelA) edge[out=0,in=180,->] node {$\push{}$} (LabelB);
  \draw (LabelB) edge[out=0,in=180,->] node {$\pop{}{x}$} (LabelC);
  \draw (0A) edge[out=0,in=180,|->] (1B);
  \draw (1A) edge[out=0,in=180,|->] (2B);
  \draw (0B) edge[out=0,in=180,|->] (4C);
  \draw (1B) edge[out=0,in=180,|->] (0C);
  \draw (2B) edge[out=0,in=180,|->] (1C);

  \node [right of=1C, node distance=1cm] {\Large $=$};

  \node [anchor=west, right of=LabelC, node distance=2cm] (LabelD) {$\Gamma$};
  \node [anchor=west, below of=LabelD] (0D) {0};
  \node [anchor=west, below of=0D] (1D) {1};
  \node [anchor=west, below of=1D] (2D) {$\vdots$};

  \node [anchor=east, right of=LabelD, node distance=2.5cm] (LabelE) {$\Gamma$};
  \node [anchor=east, below of=LabelE] (0E) {0};
  \node [anchor=east, below of=0E] (1E) {1};
  \node [anchor=east, below of=1E] (2E) {$\vdots$};

  \draw (LabelD) edge[out=0,in=180,->] node {$\id$} (LabelE);
  \draw (0D) edge[out=0,in=180,|->] (0E);
  \draw (1D) edge[out=0,in=180,|->] (1E);
\end{tikzpicture}
}
\end{center}

%% file: fig/lemma/pop-zero-after-suc-push.tex
\scalebox{0.8}{
\begin{tikzpicture}[node distance=1.5cm, auto, baseline={([yshift=-.5ex]current bounding box.center)}]
  \node (GammaA) {$\suc{\Gamma}$};
  \node (GammaB) [node distance=3cm, right of=GammaA] {$\plus{\Gamma}{2}$};
  \node (GammaC) [below of=GammaB] {$\suc{\Gamma}$};
  \draw[->] (GammaA) to node {$\suc{\push{\Gamma}}$} (GammaB);
  \draw[-,double distance=1pt] (GammaA) to node [xshift=0.8em, swap] {} (GammaC);
  \draw[->, shift left=0.2em] (GammaB) to node {$\pop{\suc{\Gamma}}{0}$} (GammaC);
\end{tikzpicture}
}

%% file: fig/lemma/illustrate/pop-zero-after-suc-push.tex
\begin{center}
\scalebox{0.8}{
\begin{tikzpicture}[node distance=0.5cm, auto]
  \node [anchor=west] at (0,0) (LabelA) {$\suc{\Gamma}$};
  \node [anchor=west, below of=LabelA] (0A) {0};
  \node [anchor=west, below of=0A] (1A) {1};
  \node [anchor=west, below of=1A] (2A) {2};
  \node [anchor=west, below of=2A] (3A) {$\vdots$};

  \node [anchor=east, right of=LabelA, node distance=2.5cm] (LabelB) {$\plus{\Gamma}{2}$};
  \node [anchor=east, below of=LabelB] (0B) {0};
  \node [anchor=east, below of=0B] (1B) {1};
  \node [anchor=east, below of=1B] (2B) {2};
  \node [anchor=east, below of=2B] (3B) {3};
  \node [anchor=east, below of=3B] (4B) {$\vdots$};

  \node [anchor=east, right of=LabelB, node distance=2.5cm] (LabelC) {$\suc{\Gamma}$};
  \node [anchor=east, below of=LabelC] (0C) {0};
  \node [anchor=east, below of=0C] (1C) {1};
  \node [anchor=east, below of=1C] (2C) {2};
  \node [anchor=east, below of=2C] (3C) {$\vdots$};

  \draw (LabelA) edge[out=0,in=180,->] node {$\suc{\push{}}$} (LabelB);
  \draw (LabelB) edge[out=0,in=180,->] node {$\pop{}{0}$} (LabelC);
  \draw (0A) edge[out=0,in=180,|->] (0B);
  \draw (1A) edge[out=0,in=180,|->] (2B);
  \draw (2A) edge[out=0,in=180,|->] (3B);
  \draw (0B) edge[out=0,in=180,|->] (0C);
  \draw (1B) edge[out=0,in=180,|->] (0C);
  \draw (2B) edge[out=0,in=180,|->] (1C);
  \draw (3B) edge[out=0,in=180,|->] (2C);

  \node [right of=1C, node distance=1cm] {\Large $=$};

  \node [anchor=west, right of=LabelC, node distance=2cm] (LabelD) {$\suc{\Gamma}$};
  \node [anchor=west, below of=LabelD] (0D) {0};
  \node [anchor=west, below of=0D] (1D) {1};
  \node [anchor=west, below of=1D] (2D) {2};
  \node [anchor=west, below of=2D] (3D) {$\vdots$};

  \node [anchor=east, right of=LabelD, node distance=2.5cm] (LabelE) {$\suc{\Gamma}$};
  \node [anchor=east, below of=LabelE] (0E) {0};
  \node [anchor=east, below of=0E] (1E) {1};
  \node [anchor=east, below of=1E] (2E) {2};
  \node [anchor=east, below of=2E] (3E) {$\vdots$};

  \draw (LabelD) edge[out=0,in=180,->] node {$\id$} (LabelE);
  \draw (0D) edge[out=0,in=180,|->] (0E);
  \draw (1D) edge[out=0,in=180,|->] (1E);
  \draw (2D) edge[out=0,in=180,|->] (2E);
\end{tikzpicture}
}
\end{center}

%% file: fig/lemma/swap-after-suc-swap-after-swap.tex
\scalebox{0.8}{
\begin{tikzpicture}[node distance=1.5cm, auto, baseline={([yshift=-.5ex]current bounding box.center)}]
  \node (GammaA) {$\plus{\Gamma}{3}$};
  \node (GammaB) [right of=GammaA, above of=GammaA] {$\plus{\Gamma}{3}$};
  \node (GammaC) [node distance=3cm, right of=GammaB] {$\plus{\Gamma}{3}$};
  \node (GammaD) [right of=GammaA, below of=GammaA] {$\plus{\Gamma}{3}$};
  \node (GammaE) [node distance=3cm, right of=GammaD] {$\plus{\Gamma}{3}$};
  \node (GammaF) [right of=GammaC, below of=GammaC] {$\plus{\Gamma}{3}$};
  \draw[->] (GammaA) to node {$\sub{\swapR}{\suc{\Gamma}}$} (GammaB);
  \draw[->] (GammaB) to node {$\suc{\sub{\swapR}{\Gamma}}$} (GammaC);
  \draw[->] (GammaA) to node [swap] {$\suc{\sub{\swapR}{\Gamma}}$} (GammaD);
  \draw[->] (GammaD) to node [swap] {$\sub{\swapR}{\suc{\Gamma}}$} (GammaE);
  \draw[->] (GammaC) to node {$\sub{\swapR}{\suc{\Gamma}}$} (GammaF);
  \draw[->] (GammaE) to node [swap] {$\suc{\sub{\swapR}{\Gamma}}$} (GammaF);
\end{tikzpicture}
}

%% file: fig/lemma/illustrate/swap-after-suc-swap-after-swap.tex
\begin{center}
\scalebox{0.8}{
\begin{tikzpicture}[node distance=0.5cm, auto]
  \node [anchor=west] at (0,0) (LabelA) {$\plus{\Gamma}{3}$};
  \node [anchor=west, below of=LabelA] (0A) {0};
  \node [anchor=west, below of=0A] (1A) {1};
  \node [anchor=west, below of=1A] (2A) {2};
  \node [anchor=west, below of=2A] (3A) {$\vdots$};

  \node [anchor=east, right of=LabelA, node distance=2.5cm] (LabelB) {$\plus{\Gamma}{3}$};
  \node [anchor=east, below of=LabelB] (0B) {0};
  \node [anchor=east, below of=0B] (1B) {1};
  \node [anchor=east, below of=1B] (2B) {2};
  \node [anchor=east, below of=2B] (3B) {$\vdots$};

  \node [anchor=east, right of=LabelB, node distance=2.5cm] (LabelC) {$\plus{\Gamma}{3}$};
  \node [anchor=east, below of=LabelC] (0C) {0};
  \node [anchor=east, below of=0C] (1C) {1};
  \node [anchor=east, below of=1C] (2C) {2};
  \node [anchor=east, below of=2C] (3C) {$\vdots$};

  \node [anchor=west, right of=LabelC, node distance=2.5cm] (LabelD) {$\plus{\Gamma}{3}$};
  \node [anchor=west, below of=LabelD] (0D) {0};
  \node [anchor=west, below of=0D] (1D) {1};
  \node [anchor=west, below of=1D] (2D) {2};
  \node [anchor=west, below of=2D] (3D) {$\vdots$};

  \draw (LabelA) edge[out=0,in=180,->] node {$\suc{\swap{\Gamma}}$} (LabelB);
  \draw (LabelB) edge[out=0,in=180,->] node {$\swap{\suc{\Gamma}}$} (LabelC);
  \draw (LabelC) edge[out=0,in=180,->] node {$\suc{\swap{\Gamma}}$} (LabelD);
  \draw (0A) edge[,out=0,in=180,|->] (1B);
  \draw (1A) edge[out=0,in=180,|->] (0B);
  \draw (2A) edge[,out=0,in=180,|->] (2B);
  \draw (0B) edge[out=0,in=180,|->] (0C);
  \draw (1B) edge[,out=0,in=180,|->] (2C);
  \draw (2B) edge[,out=0,in=180,|->] (1C);
  \draw (0C) edge[out=0,in=180,|->] (1D);
  \draw (1C) edge[,out=0,in=180,|->] (0D);
  \draw (2C) edge[,out=0,in=180,|->] (2D);

  \node [right of=2B, below of=0B, yshift=-0.5cm, node distance=1.25cm] {\Large $=$};

  \node [anchor=west] at (0,-3.25) (LabelAA) {$\plus{\Gamma}{3}$};
  \node [anchor=west, below of=LabelAA] (0AA) {0};
  \node [anchor=west, below of=0AA] (1AA) {1};
  \node [anchor=west, below of=1AA] (2AA) {2};
  \node [anchor=west, below of=2AA] (3AA) {$\vdots$};

  \node [anchor=east, right of=LabelAA, node distance=2.5cm] (LabelBB) {$\plus{\Gamma}{3}$};
  \node [anchor=east, below of=LabelBB] (0BB) {0};
  \node [anchor=east, below of=0BB] (1BB) {1};
  \node [anchor=east, below of=1BB] (2BB) {2};
  \node [anchor=east, below of=2BB] (3BB) {$\vdots$};

  \node [anchor=east, right of=LabelBB, node distance=2.5cm] (LabelCC) {$\plus{\Gamma}{3}$};
  \node [anchor=east, below of=LabelCC] (0CC) {0};
  \node [anchor=east, below of=0CC] (1CC) {1};
  \node [anchor=east, below of=1CC] (2CC) {2};
  \node [anchor=east, below of=2CC] (3CC) {$\vdots$};

  \node [anchor=west, right of=LabelCC, node distance=2.5cm] (LabelDD) {$\plus{\Gamma}{3}$};
  \node [anchor=west, below of=LabelDD] (0DD) {0};
  \node [anchor=west, below of=0DD] (1DD) {1};
  \node [anchor=west, below of=1DD] (2DD) {2};
  \node [anchor=west, below of=2DD] (3DD) {$\vdots$};

  \draw (LabelAA) edge[out=0,in=180,->] node {$\swap{\suc{\Gamma}}$} (LabelBB);
  \draw (LabelBB) edge[out=0,in=180,->] node {$\suc{\swap{\Gamma}}$} (LabelCC);
  \draw (LabelCC) edge[out=0,in=180,->] node {$\swap{\suc{\Gamma}}$} (LabelDD);
  \draw (0AA) edge[,out=0,in=180,|->] (0BB);
  \draw (1AA) edge[out=0,in=180,|->] (2BB);
  \draw (2AA) edge[,out=0,in=180,|->] (1BB);
  \draw (0BB) edge[,out=0,in=180,|->] (1CC);
  \draw (1BB) edge[,out=0,in=180,|->] (0CC);
  \draw (2BB) edge[out=0,in=180,|->] (2CC);
  \draw (0CC) edge[,out=0,in=180,|->] (0DD);
  \draw (1CC) edge[,out=0,in=180,|->] (2DD);
  \draw (2CC) edge[out=0,in=180,|->] (1DD);
\end{tikzpicture}
}
\end{center}

%% file: fig/lemma/pop-swap.tex
\scalebox{0.8}{
\begin{tikzpicture}[node distance=3cm, auto, baseline={([yshift=-.5ex]current bounding box.center)}]
  \node (P) {$\plus{\Gamma}{2}$};
  \node (PPrime) [right of=P] {$\plus{\Gamma}{2}$};
  \node (Q) [right of=PPrime] {$\suc{\Gamma}$};
  \draw[->, shift left=0.2em] (P) to node {$\swapR$} (PPrime);
  \draw[->, shift right=0.2em] (P) to node [swap] {$\id$} (PPrime);
  \draw[->] (PPrime) to node {$\pop{\suc{\Gamma}}{0}$} (Q);
\end{tikzpicture}
}

%% file: fig/lemma/illustrate/pop-swap.tex
\begin{center}
\scalebox{0.8}{
\begin{tikzpicture}[node distance=0.5cm, auto]
  \node [anchor=west] at (0,0) (LabelA) {$\plus{\Gamma}{2}$};
  \node [anchor=west, below of=LabelA] (0A) {0};
  \node [anchor=west, below of=0A] (1A) {1};
  \node [anchor=west, below of=1A] (2A) {2};
  \node [anchor=west, below of=2A] (3A) {$\vdots$};

  \node [anchor=east, right of=LabelA, node distance=2.5cm] (LabelB) {$\plus{\Gamma}{2}$};
  \node [anchor=east, below of=LabelB] (0B) {0};
  \node [anchor=east, below of=0B] (1B) {1};
  \node [anchor=east, below of=1B] (2B) {2};
  \node [anchor=east, below of=2B] (3B) {$\vdots$};

  \node [anchor=east, right of=LabelB, node distance=2.5cm] (LabelC) {$\suc{\Gamma}$};
  \node [anchor=east, below of=LabelC] (0C) {0};
  \node [anchor=east, below of=0C] (1C) {1};
  \node [anchor=east, below of=1C] (2C) {$\vdots$};

  \draw (LabelA) edge[out=0,in=180,->] node {$\swap{}$} (LabelB);
  \draw (LabelB) edge[out=0,in=180,->] node {$\pop{}{0}$} (LabelC);
  \draw (0A) edge[out=0,in=180,|->] (1B);
  \draw (1A) edge[out=0,in=180,|->] (0B);
  \draw (2A) edge[out=0,in=180,|->] (2B);
  \draw (0B) edge[out=0,in=180,|->] (0C);
  \draw (1B) edge[out=0,in=180,|->] (0C);
  \draw (2B) edge[out=0,in=180,|->] (1C);

  \node [right of=1C, node distance=1cm] {\Large $=$};

  \node [anchor=west, right of=LabelC, node distance=2cm] (LabelD) {$\plus{\Gamma}{2}$};
  \node [anchor=west, below of=LabelD] (0D) {0};
  \node [anchor=west, below of=0D] (1D) {1};
  \node [anchor=west, below of=1D] (2D) {2};
  \node [anchor=west, below of=1D] (3D) {$\vdots$};

  \node [anchor=east, right of=LabelD, node distance=2.5cm] (LabelE) {$\suc{\Gamma}$};
  \node [anchor=east, below of=LabelE] (0E) {0};
  \node [anchor=east, below of=0E] (1E) {1};
  \node [anchor=east, below of=1E] (2E) {$\vdots$};

  \draw (LabelD) edge[out=0,in=180,->] node {$\pop{}{0}$} (LabelE);
  \draw (0D) edge[out=0,in=180,|->] (0E);
  \draw (1D) edge[out=0,in=180,|->] (0E);
  \draw (2D) edge[out=0,in=180,|->] (1E);
\end{tikzpicture}
}
\end{center}

%% file: fig/lemma/swap-push.tex
\scalebox{0.8}{
\begin{tikzpicture}[node distance=1.5cm, auto, baseline={([yshift=-.5ex]current bounding box.center)}]
  \node (P) {$\suc{\Gamma}$};
  \node (PPrime) [node distance=3cm, right of=P] {$\plus{\Gamma}{2}$};
  \node (Q) [below of=PPrime] {$\plus{\Gamma}{2}$};
  \draw[->] (P) to node {$\suc{\push{\Gamma}}$} (PPrime);
  \draw[->] (P) to node [xshift=0.8em, swap] {$\push{\suc{\Gamma}}$} (Q);
  \draw[->, shift left=0.2em] (PPrime) to node {$\swapR$} (Q);
  \draw[->, shift left=0.2em] (Q) to node {$\swapR$} (PPrime);
\end{tikzpicture}
}

%% file: fig/lemma/illustrate/swap-push-1.tex
\begin{center}
\scalebox{0.8}{
\begin{tikzpicture}[node distance=0.5cm, auto]
  \node [anchor=west] at (0,0) (LabelA) {$\suc{\Gamma}$};
  \node [anchor=west, below of=LabelA] (0A) {0};
  \node [anchor=west, below of=0A] (1A) {1};
  \node [anchor=west, below of=1A] (2A) {$\vdots$};

  \node [anchor=east, right of=LabelA, node distance=2.5cm] (LabelB) {$\plus{\Gamma}{2}$};
  \node [anchor=east, below of=LabelB] (0B) {0};
  \node [anchor=east, below of=0B] (1B) {1};
  \node [anchor=east, below of=1B] (2B) {2};
  \node [anchor=east, below of=2B] (3B) {$\vdots$};

  \node [anchor=east, right of=LabelB, node distance=2.5cm] (LabelC) {$\plus{\Gamma}{2}$};
  \node [anchor=east, below of=LabelC] (0C) {0};
  \node [anchor=east, below of=0C] (1C) {1};
  \node [anchor=east, below of=1C] (2C) {2};
  \node [anchor=east, below of=2C] (3C) {$\vdots$};

  \draw (LabelA) edge[out=0,in=180,->] node {$\push{\suc{\Gamma}}$} (LabelB);
  \draw (LabelB) edge[out=0,in=180,->] node {$\swap{\Gamma}$} (LabelC);
  \draw (0A) edge[out=0,in=180,|->] (1B);
  \draw (1A) edge[out=0,in=180,|->] (2B);
  \draw (0B) edge[out=0,in=180,|->] (1C);
  \draw (1B) edge[out=0,in=180,|->] (0C);
  \draw (2B) edge[out=0,in=180,|->] (2C);

  \node [right of=1C, node distance=1cm] {\Large $=$};

  \node [anchor=west, right of=LabelC, node distance=2cm] (LabelD) {$\suc{\Gamma}$};
  \node [anchor=west, below of=LabelD] (0D) {0};
  \node [anchor=west, below of=0D] (1D) {1};
  \node [anchor=west, below of=1D] (2D) {$\vdots$};

  \node [anchor=east, right of=LabelD, node distance=2.5cm] (LabelE) {$\plus{\Gamma}{2}$};
  \node [anchor=east, below of=LabelE] (0E) {0};
  \node [anchor=east, below of=0E] (1E) {1};
  \node [anchor=east, below of=1E] (2E) {2};
  \node [anchor=east, below of=2E] (3E) {$\vdots$};

  \draw (LabelD) edge[out=0,in=180,->] node {$\suc{\push{\Gamma}}$} (LabelE);
  \draw (0D) edge[out=0,in=180,|->] (0E);
  \draw (1D) edge[out=0,in=180,|->] (2E);
\end{tikzpicture}
}
\end{center}

%% file: fig/lemma/illustrate/swap-push-2.tex
\begin{center}
\scalebox{0.8}{
\begin{tikzpicture}[node distance=0.5cm, auto]
  \node [anchor=west] at (0,0) (LabelA) {$\suc{\Gamma}$};
  \node [anchor=west, below of=LabelA] (0A) {0};
  \node [anchor=west, below of=0A] (1A) {1};
  \node [anchor=west, below of=1A] (2A) {$\vdots$};

  \node [anchor=east, right of=LabelA, node distance=2.5cm] (LabelB) {$\plus{\Gamma}{2}$};
  \node [anchor=east, below of=LabelB] (0B) {0};
  \node [anchor=east, below of=0B] (1B) {1};
  \node [anchor=east, below of=1B] (2B) {2};
  \node [anchor=east, below of=2B] (3B) {$\vdots$};

  \node [anchor=east, right of=LabelB, node distance=2.5cm] (LabelC) {$\plus{\Gamma}{2}$};
  \node [anchor=east, below of=LabelC] (0C) {0};
  \node [anchor=east, below of=0C] (1C) {1};
  \node [anchor=east, below of=1C] (2C) {2};
  \node [anchor=east, below of=2C] (3C) {$\vdots$};

  \draw (LabelA) edge[out=0,in=180,->] node {$\suc{\push{\Gamma}}$} (LabelB);
  \draw (LabelB) edge[out=0,in=180,->] node {$\swap{\Gamma}$} (LabelC);
  \draw (0A) edge[out=0,in=180,|->] (0B);
  \draw (1A) edge[out=0,in=180,|->] (2B);
  \draw (0B) edge[out=0,in=180,|->] (1C);
  \draw (1B) edge[out=0,in=180,|->] (0C);
  \draw (2B) edge[out=0,in=180,|->] (2C);

  \node [right of=1C, node distance=1cm] {\Large $=$};

  \node [anchor=west, right of=LabelC, node distance=2cm] (LabelD) {$\suc{\Gamma}$};
  \node [anchor=west, below of=LabelD] (0D) {0};
  \node [anchor=west, below of=0D] (1D) {1};
  \node [anchor=west, below of=1D] (2D) {$\vdots$};

  \node [anchor=east, right of=LabelD, node distance=2.5cm] (LabelE) {$\plus{\Gamma}{2}$};
  \node [anchor=east, below of=LabelE] (0E) {0};
  \node [anchor=east, below of=0E] (1E) {1};
  \node [anchor=east, below of=1E] (2E) {2};
  \node [anchor=east, below of=2E] (3E) {$\vdots$};

  \draw (LabelD) edge[out=0,in=180,->] node {$\push{\suc{\Gamma}}$} (LabelE);
  \draw (0D) edge[out=0,in=180,|->] (1E);
  \draw (1D) edge[out=0,in=180,|->] (2E);
\end{tikzpicture}
}
\end{center}

%% file: fig/lemma/push-comm-pop-comm.tex
\scalebox{0.8}{
\begin{tikzpicture}[node distance=1.5cm, auto]
  \node (P) {$\Gamma$};
  \node (Q) [below of=P] {$\Delta$};
  \node (PPrime) [node distance=2.6cm, right of=P] {$\suc{\Gamma}$};
  \node (QPrime) [below of=PPrime] {$\suc{\Delta}$};
  \node (PDoublePrime) [node distance=2.9cm, right of=PPrime] {$\Gamma$};
  \node (QDoublePrime) [below of=PDoublePrime] {$\Delta$};
  \draw[->] (P) to node {$\push{\Gamma}$} (PPrime);
  \draw[->] (Q) to node [swap] {$\push{\Delta}$} (QPrime);
  \draw[->] (QPrime) to node [swap] {$\pop{\Delta}{\rho x}$} (QDoublePrime);
  \draw[->] (P) to node [swap] {$\rho$} (Q);
  \draw[->] (PPrime) to node {$\suc{\rho}$} (QPrime);
  \draw[->] (PPrime) to node {$\pop{\Gamma}{x}$} (PDoublePrime);
  \draw[->] (PDoublePrime) to node {$\rho$} (QDoublePrime);
\end{tikzpicture}
}

%% file: fig/lemma/swap-suc-suc.tex
\scalebox{0.8}{
\begin{tikzpicture}[node distance=1.5cm, auto]
  \node (P) {$\plus{\Gamma}{2}$};
  \node (Q) [below of=P] {$\plus{\Delta}{2}$};
  \node (PPrime) [node distance=2.8cm, right of=P] {$\plus{\Gamma}{2}$};
  \node (QPrime) [below of=PPrime] {$\plus{\Delta}{2}$};
  \draw[->] (P) to node {$\swap{\Gamma}$} (PPrime);
  \draw[->] (Q) to node [swap] {$\swap{\Delta}$} (QPrime);
  \draw[->] (P) to node [swap] {$\plus{\rho}{2}$} (Q);
  \draw[->] (PPrime) to node {$\plus{\rho}{2}$} (QPrime);
\end{tikzpicture}
}

%% file: sec/appendix/additional-proofs.tex
\section{Additional proofs}
\label{app:additional-proofs}

\paragraph{Proof of \lemref{concurrent-transitions:renaming:transition}.}
By the following mutually recursive proofs-by-induction on the
derivations. The various renaming lemmas needed to enable the induction
hypothesis in each case are omitted.

\vspace{8pt}
\input{fig/renaming/transition}

\subsection{Additional illustrative cases of \thmref{concurrent-transitions:cofinality}}
\label{app:additional-proofs:concurrent-transitions:cofinality}

\paragraph{Example: permuting concurrent extrusions (different binders).}

First, note that the residuals of bound output transitions are not
themselves necessarily bound. More specifically, the residuals of the
output transition on $\piBoundOutput{x}$ with the output on
$\piBoundOutput{z}$ is bound only if the outputs represent extrusions of
different $\nu$-binders. In this section we consider only the case when
the concurrent extrusions are of different $\nu$-binders.

In this case, each binder is unaffected by the extrusion of the other,
and the residuals remain bound outputs, shifted into $\suc{\Gamma}$ as
usual. The general form of such residuals is:

\input{fig/residual/illustrate/capture-different-binder-2}

\noindent where $\phi$ ranges over braiding congruence. Then the residual
is able to handle the inner extrusion, with the resulting $\tau$ action
again propagated through the outer binder:

\input{fig/residual/illustrate/capture-different-binder-2a}

\input{fig/residual/illustrate/capture-different-binder-3}

\paragraph{Example: permuting concurrent extrusions (same binder).}

Consider the process
$\piRestrict{(\piPar{\piAction{\piOutput{\suc{x}}{0}}{P}}{\piAction{\piOutput{\suc{z}}{0}}{Q}})}$,
as described in Cristescu \etal.~\cite{cristescu13}. There are two
concurrent outputs, both of which try to extrude the top-level binder.
Suppose we take the $\piOutput{\suc{x}}{0}$ action first:

\input{fig/residual/illustrate/extrude-0}

If we then take the $\piOutput{\suc{z}}{0}$ action, the enclosing
$\nu$-binder no longer exists, and so $\piOutput{\suc{z}}{0}$ simply
propagates as a non-bound action.

\input{fig/residual/illustrate/extrude-1}

\paragraph{Example: permuting one extrusion-rendezvous with another.}

Now consider what happens when the extrusions from the previous example
eventually rendezvous with a compatible input.

{\small
\begin{align*}
\piParLNu{E}{F}: \Gamma \vdash \piPar{P}{Q}
&\transition{\tau}
\Gamma \vdash \piRestrict{(\piPar{R}{S})}
\\
\piParLNu{E'}{F'}: \Gamma \vdash \piPar{P}{Q}
&\transition{\tau}
\Gamma \vdash \piRestrict{(\piPar{R'}{S'})}
\end{align*}
}

\input{fig/residual/illustrate/capture-same-binder-1}

\noindent When the extrusions are of the same $\nu$-binder, and the
residual outputs are not bound, then we have:

\input{fig/residual/illustrate/capture-same-binder-2}

\noindent and the residual of one extrusion-handling after another is a
plain communication, with the resulting $\tau$ action simply propagated
through the second $\nu$ binder:

\vspace{-5pt}
\input{fig/residual/illustrate/capture-same-binder-3}

\noindent Here $\alpha$ is the equality $(\pop{}{0}) \after \swapR =
\pop{}{0}$ (\lemref{pop-swap}) applied to $P'$.

\paragraph{Example: permuting bound actions propagating through a binder.}

Now suppose we have a process of the form $\piRestrict{P}$ which has two
concurrent transitions propagating an input action through the $\nu$
binder:

\input{fig/residual/illustrate/propagate-1}

\noindent (The derivations are valid because both $\piInput{\suc{x}}$
and $\piInput{\suc{z}}$ are of the form $\ren{\pushR}{b}$.) The
residuals of $E$ and $E'$ with respect to each other have the form:

\input{fig/residual/illustrate/propagate-2}

\noindent We can use these residuals to define the following composite
residual
$\residual{(\piRestrictA{\piInput{u}}{E'})}{(\piRestrictA{\piInput{x}}{E})}$:

\input{fig/residual/illustrate/propagate-3}

\noindent noting that $\ren{\swapR}{(\piInput{\plus{u}{2}})} =
\piInput{\plus{u}{2}}$ by \lemref{swap-suc-suc}. The complementary
residual
$\residual{(\piRestrictA{\piInput{x}}{E})}{(\piRestrictA{\piInput{u}}{E'})}$
is similar, with $x$ instead of $u$ and $R'$ instead of $R$. It remains
to show that the terminal states are $\swapR$-congruent:

\vspace{-10pt}
\input{fig/residual/illustrate/propagate-4}

\paragraph{Example: permuting extruding rendezvous and unhandled extrusion.}

Of course concurrent transitions are not always as symmetric as the ones
we have seen. Here a name extrusion which has a successful rendezvous,
resulting in a $\tau$ action, is concurrent with another which does not
and which therefore propagates as a bound output:

\vspace{-5pt}
{\small
\begin{align*}
\piParR{\piBoundOutput{u}}{P}{F}: \Gamma \vdash \piPar{P}{Q}
&\transition{\piBoundOutput{u}}
\suc{\Gamma} \vdash \piPar{\ren{\push{}}{P}}{S}
\\
\piParLNu{E}{F'}: \Gamma \vdash \piPar{P}{Q}
&\transition{\tau}
\Gamma \vdash \piRestrict{(\piPar{R}{S'})}
\end{align*}
}
As before, it matters whether the extrusions
$\cxtRaw{F}{\piBoundOutput{x}} \concur
\cxtRaw{F'}{\piBoundOutput{u}}$ are of the same or different
binders.

\emph{Sub-case: extrusions of same binders.} In this case, the residuals
$\residual{F'}{F}$ and $\residual{F}{F'}$ become sends of index 0, the
binder being extruded.

\input{fig/residual/illustrate/extrude-propagate-1}

\noindent For the other residual, we can derive:

\input{fig/residual/illustrate/extrude-propagate-2}

\noindent with $Q' \congEq Q''$, and noting that $\pop{}{0}$ retracts
$\suc{\push{}}$ (\lemref{pop-zero-after-suc-push} below).

\emph{Sub-case: extrusions of different binders.} In this case the
residuals $\residual{F'}{F}$ and $\residual{F}{F'}$ remain bound
outputs. Then, with the $\ren{\push{}}{E}$ derivation as before, we can
derive:

\input{fig/residual/illustrate/extrude-propagate-3}

\noindent and for the other residual:

\input{fig/residual/illustrate/extrude-propagate-4}

\noindent with $\ren{\swapR}{Q'} \congEq Q''$. It remains to establish a
$\congEq$-path between the two terminal processes. We have $Q' \congEq
\ren{\swapR}{Q''}$ by functionality and involutivity of $\swapR$, and
$\suc{\push{}} = \swapR \after \push{}$ by \lemref{swap-push} and then
the rest follows by reflexivity and congruence.

\subsection{Cofinality for \thmref{causal-equivalence:residual:braiding-congruence}}
\label{app:additional-proofs:causal-equivalence:residual:braiding-congruence}

\input{fig/residual/congruence/nu-nu-swap-cases}

\figref{residual:congruence:nu-nu-swap-cases} illustrates cofinality for
the $\congRestrictSwap{}$ cases, omitting the renaming lemmas used as
type-level coercions. The $\congRestrictSwapInv{}$ cases are symmetric.

%% file: fig/renaming/transition.tex
\begin{minipage}[t]{0.49\linewidth}
\noindent \shadebox{$\ren{\rho}{\cxtRaw{E}{c}}$}
{\small
\begin{align*}
\ren{\rho}{(\piAction{\piOutput{x}{y}}{P})}
&=
\piAction{\piOutput{\rho x}{\rho y}}{\ren{\rho}{P}}
\\
\ren{\rho}{(\piChoice{E}{F})}
&=
\piChoice{\ren{\rho}{E}}{\ren{\rho}{F}}
\\
\ren{\rho}{(\piParR{c}{P}{F})}
&=
\piParR{\ren{\rho}{c}}{\ren{\rho}{P}}{\ren{\rho}{F}}
\\
\ren{\rho}{(\piParL{c}{E}{Q})}
&=
\piParL{\ren{\rho}{c}}{\ren{\rho}{E}}{\ren{\rho}{Q}}
\\
\ren{\rho}{(\piParLTau{E}{F}{y})}
&=
\piParLTau{\ren{\rho}{E}}{\ren{\rho}{F}}{\ren{\rho}{y}}
\\
\ren{\rho}{(\piParLNu{E}{F})}
&=
\piParLNu{\ren{\rho}{E}}{\ren{\rho}{F}}
\\
\ren{\rho}{(\piRestrictA{c}{E})}
&=
\piRestrictA{\ren{\rho}{c}}{\ren{(\suc{\rho})}{E}}
\\
\ren{\rho}{(\piReplicate{E})}
&=
\piReplicate{\ren{\rho}{E}}
\end{align*}
}
\end{minipage}
\begin{minipage}[t]{0.49\linewidth}
\noindent \shadebox{$\ren{\rho}{\cxtRaw{E}{b}}$}
{\small
\begin{align*}
\ren{\rho}{(\piAction{\piInput{x}}{P})}
&=
\piAction{\piInput{\rho x}}{\ren{(\suc{\rho})}{P}}
\\
\ren{\rho}{(\piChoice{E}{F})}
&=
\piChoice{\ren{\rho}{E}}{\ren{\rho}{F}}
\\
\ren{\rho}{(\piParR{b}{P}{F})}
&=
\piParR{\ren{\rho}{b}}{\ren{\rho}{P}}{\ren{\rho}{F}}
\\
\ren{\rho}{(\piParL{b}{E}{Q})}
&=
\piParL{\ren{\rho}{b}}{\ren{\rho}{E}}{\ren{\rho}{Q}}
\\
\ren{\rho}{(\piRestrictOutput{E})}
&=
\piRestrictOutput{\ren{(\suc{\rho})}{E}}
\\
\ren{\rho}{(\piRestrictA{b}{E})}
&=
\piRestrictA{\ren{\rho}{b}}{\ren{(\suc{\rho})}{E}}
\\
\ren{\rho}{(\piReplicate{E})}
&=
\piReplicate{\ren{\rho}{E}}
\end{align*}
}
\end{minipage}

%% file: fig/residual/illustrate/capture-different-binder-2.tex
\begin{center}
\scalebox{0.8}{
\begin{tikzpicture}[node distance=1.5cm, auto]
  \node (Q) [node distance=2cm] {$\Gamma \vdash Q$};
  \node (SPrime) [below of=Q, right of=Q] {$\suc{\Gamma} \vdash S'$};
  \node (S) [right of=Q, above of=Q] {$\suc{\Gamma} \vdash S$};
  \node (QPrime) [node distance=3.5cm, right of=S] {$\plus{\Gamma}{2} \vdash Q'$};
  \node (SwapQPrime) [below of=QPrime] {$\plus{\Gamma}{2} \vdash \ren{\swapR}{Q'}$};
  \node (QDoublePrime) [node distance=3.5cm, right of=SPrime] {$\plus{\Gamma}{2} \vdash Q''$};
  \draw[->] (Q) to node [yshift=-1ex] {$\cxtRaw{F}{\piBoundOutput{x}}$} (S);
  \draw[->] (Q) to node [yshift=1ex,swap] {$\cxtRaw{F'}{\piBoundOutput{u}}$} (SPrime);
  \draw[->] (S) to node {$\cxtRaw{(\residual{F'}{F})}{\piBoundOutput{u+1}}$} (QPrime);
  \draw[->] (SPrime) to node [swap] {$\cxtRaw{(\residual{F}{F'})}{\piBoundOutput{x+1}}$} (QDoublePrime);
  \draw[->] (QPrime) to node {$\ren{\swapR}{}$} (SwapQPrime);
  \draw[->] (SwapQPrime) to node {$\phi$} (QDoublePrime);
\end{tikzpicture}
}
\end{center}

%% file: fig/residual/illustrate/capture-different-binder-2a.tex
\begin{center}
\scalebox{\smathparscale}{
\begin{smathpar}
\inferrule*[left={\ruleName{$\piParLNu{\param}{\param}$}}]
{
  \inferrule*[left={\ruleName{$E$}}]
  {
    \vdots
  }
  {
    \Gamma \vdash P
    \transitionWithoutSmash{\piInput{x}}
    R
  }
  \\
  \inferrule*[left={\ruleName{$F$}}]
  {
    \vdots
  }
  {
    \Gamma \vdash Q
    \transitionWithoutSmash{\piBoundOutput{x}}
    S
  }
}
{
  \Gamma \vdash \piPar{P}{Q}
  \transitionWithoutSmash{\tau}
  \piRestrict{(\piPar{R}{S})}
}
\end{smathpar}
}
\end{center}

%% file: fig/residual/illustrate/capture-different-binder-3.tex
\begin{center}
\scalebox{0.8}{
\begin{tikzpicture}[node distance=1.5cm, auto]
  \node (PQ) [node distance=2cm] {$\Gamma \vdash \piPar{P}{Q}$};
  \node (NuRPrimeSPrime) [below of=PQ, right of=PQ] {$\Gamma \vdash \piRestrict{(\piPar{R'}{S'})}$};
  \node (RS) [right of=PQ, above of=PQ] {$\Gamma \vdash \piRestrict{(\piPar{R}{S})}$};
  \node (NuSwapPPrimeQPrime) [node distance=6cm, right of=RS] {
    $\Gamma \vdash \piRestrict{\piRestrict{(\piPar{\ren{\swapR}{P'}}{\ren{\swapR}{Q'}})}}$
  };
  \node (NuPPrimeQPrime) [below of=NuSwapPPrimeQPrime] {
    $\Gamma \vdash \piRestrict{\piRestrict{(\piPar{P'}{Q'})}}$
  };
  \node (NuPDoublePrimeQDoublePrime) [node distance=6cm, right of=NuRPrimeSPrime] {
    $\Gamma \vdash \piRestrict{\piRestrict{(\piPar{P''}{Q''})}}$
  };
  \draw[->] (PQ) to node [xshift=1ex,yshift=-1.5ex] {$\piParLNu{E}{F}$} (RS);
  \draw[->] (PQ) to node [yshift=1ex,swap] {$\piParLNu{E'}{F'}$} (NuRPrimeSPrime);
  \draw[->] (RS) to node {
    $\piRestrictA{\piTau}{(\piParLNu{\residual{E'}{E}}{\residual{F'}{F}})}$
  } (NuSwapPPrimeQPrime);
  \draw[->] (NuRPrimeSPrime) to node [swap] {
    $\piRestrictA{\piTau}{(\piParLNu{\residual{E}{E'}}{\residual{F}{F'}})}$
  } (NuPDoublePrimeQDoublePrime);
  \draw[->] (NuSwapPPrimeQPrime) to node {$\congRestrictSwap{\piPar{P'}{Q'}}$} (NuPPrimeQPrime);
  \draw[->] (NuPPrimeQPrime) to node {$\piRestrict{\piRestrict{(\piPar{\phi}{\psi}})}$} (NuPDoublePrimeQDoublePrime);
\end{tikzpicture}
}
\end{center}

%% file: fig/residual/illustrate/extrude-0.tex
\begin{center}
\scalebox{\smathparscale}{
\begin{smathpar}
\inferrule*[left={\ruleName{$\piRestrictOutput{\param}$}}]
{
  \inferrule*[left={\ruleName{$\piParL{\param}{\param}{\param}$}}]
  {
    \piAction{\piOutput{\suc{x}}{0}}{P}
    \transitionWithoutSmash{\piOutput{\suc{x}}{0}}
    P
  }
  {
    \suc{\Gamma} \vdash \piPar{\piAction{\piOutput{\suc{x}}{0}}{P}}{\piAction{\piOutput{\suc{z}}{0}}{Q}}
    \transitionWithoutSmash{\piOutput{\suc{x}}{0}}
    \suc{\Gamma} \vdash \piPar{P}{\piAction{\piOutput{\suc{z}}{0}}{Q}}
  }
}
{
  \Gamma \vdash \piRestrict{(\piPar{\piAction{\piOutput{\suc{x}}{0}}{P}}{\piAction{\piOutput{\suc{z}}{0}}{Q}})}
  \transitionWithoutSmash{\piBoundOutput{x}}
  \suc{\Gamma} \vdash \piPar{P}{\piAction{\piOutput{\suc{z}}{0}}{Q}}
}
\end{smathpar}
}
\end{center}

%% file: fig/residual/illustrate/extrude-1.tex
\begin{center}
\scalebox{\smathparscale}{
\begin{smathpar}
\inferrule*[left={\ruleName{$\piParR{\piOutput{\suc{z}}{0}}{P}{\param}$}}]
{
  \suc{\Gamma} \vdash \piAction{\piOutput{\suc{z}}{0}}{Q}
  \transitionWithoutSmash{\piOutput{\suc{z}}{0}}
  \suc{\Gamma} \vdash Q
}
{
  \suc{\Gamma} \vdash \piPar{P}{\piAction{\piOutput{\suc{z}}{0}}{Q}}
  \transitionWithoutSmash{\piOutput{\suc{z}}{0}}
  \suc{\Gamma} \vdash \piPar{P}{Q}
}
\end{smathpar}
}
\end{center}

%% file: fig/residual/illustrate/capture-same-binder-1.tex
\begin{center}
\scalebox{0.8}{
\begin{tikzpicture}[node distance=1.5cm, auto]
  \node (P) [node distance=2cm] {$\Gamma \vdash P$};
  \node (RPrime) [below of=P, right of=P] {$\suc{\Gamma} \vdash R'$};
  \node (R) [right of=P, above of=P] {$\suc{\Gamma} \vdash R$};
  \node (PPrime) [node distance=3.5cm, right of=R] {$\plus{\Gamma}{2} \vdash P'$};
  \node (SwapPPrime) [below of=PPrime] {$\plus{\Gamma}{2} \vdash \ren{\swapR}{P'}$};
  \node (PDoublePrime) [node distance=3.5cm, right of=RPrime] {$\plus{\Gamma}{2} \vdash P''$};
  \draw[->] (P) to node [yshift=-1ex] {$\cxtRaw{E}{\piInput{x}}$} (R);
  \draw[->] (P) to node [yshift=1ex,swap] {$\cxtRaw{E'}{\piInput{u}}$} (RPrime);
  \draw[->] (R) to node {$\cxtRaw{(\residual{E'}{E})}{\piInput{u+1}}$} (PPrime);
  \draw[->] (RPrime) to node [swap] {$\cxtRaw{(\residual{E}{E'})}{\piInput{x+1}}$} (PDoublePrime);
  \draw[->] (PPrime) to node {$\ren{\swapR}{}$} (SwapPPrime);
  \draw[->] (SwapPPrime) to node {$\phi$} (PDoublePrime);
\end{tikzpicture}
}
\end{center}

%% file: fig/residual/illustrate/capture-same-binder-2.tex
\begin{center}
\scalebox{0.8}{
\begin{tikzpicture}[node distance=1.5cm, auto]
  \node (Q) [node distance=2cm] {$\Gamma \vdash Q$};
  \node (SPrime) [below of=Q, right of=Q] {$\suc{\Gamma} \vdash S'$};
  \node (S) [right of=Q, above of=Q] {$\suc{\Gamma} \vdash S$};
  \node (QPrime) [node distance=3.5cm, right of=S] {$\suc{\Gamma} \vdash Q'$};
  \node (QDoublePrime) [node distance=3.5cm, right of=SPrime] {$\suc{\Gamma} \vdash Q''$};
  \draw[->] (Q) to node [yshift=-1ex] {$\cxtRaw{F}{\piBoundOutput{x}}$} (S);
  \draw[->] (Q) to node [yshift=1ex,swap] {$\cxtRaw{F'}{\piBoundOutput{u}}$} (SPrime);
  \draw[->] (S) to node {$\cxtRaw{(\residual{F'}{F})}{\piOutput{u+1}{0}}$} (QPrime);
  \draw[->] (SPrime) to node [swap] {$\cxtRaw{(\residual{F}{F'})}{\piOutput{x+1}{0}}$} (QDoublePrime);
  \draw[->] (QPrime) to node {$\psi$} (QDoublePrime);
\end{tikzpicture}
}
\end{center}

%% file: fig/residual/illustrate/capture-same-binder-3.tex
\begin{center}
\scalebox{0.8}{
\begin{tikzpicture}[node distance=1.5cm, auto]
  \node (PQ) [node distance=2cm] {$\Gamma \vdash \piPar{P}{Q}$};
  \node (NuRPrimeSPrime) [below of=PQ, right of=PQ] {$\Gamma \vdash \piRestrict{(\piPar{R'}{S'})}$};
  \node (RS) [right of=PQ, above of=PQ] {$\Gamma \vdash \piRestrict{(\piPar{R}{S})}$};
  \node (NuSwapPPrimeQPrime) [node distance=6cm, right of=RS] {
    $\Gamma \vdash \piRestrict{(\piPar{\ren{(\pop{}{0})}{\ren{\swapR}{P'}}}{Q'})}$
  };
  \node (NuPPrimeQPrime) [below of=NuSwapPPrimeQPrime] {
    $\Gamma \vdash \piRestrict{(\piPar{\ren{(\pop{}{0})}{P'}}{Q'})}$
  };
  \node (NuPDoublePrimeQDoublePrime) [node distance=6cm, right of=NuRPrimeSPrime] {
    $\Gamma \vdash \piRestrict{(\piPar{\ren{(\pop{}{0})}{P''}}{Q''})}$
  };
  \draw[->] (PQ) to node [xshift=1ex,yshift=-1.5ex] {$\piParLNu{E}{F}$} (RS);
  \draw[->] (PQ) to node [yshift=1ex,swap] {$\piParLNu{E'}{F'}$} (NuRPrimeSPrime);
  \draw[->] (RS) to node {
    $\piRestrictA{\piTau}{(\piParLTau{\residual{E'}{E}}{\residual{F'}{F}}{0})}$
  } (NuSwapPPrimeQPrime);
  \draw[->] (NuRPrimeSPrime) to node [swap] {
    $\piRestrictA{\piTau}{(\piParLTau{\residual{E}{E'}}{\residual{F}{F'}}{0})}$
  } (NuPDoublePrimeQDoublePrime);
  \draw[-,double distance=1pt] (NuSwapPPrimeQPrime) to node {$\piRestrict{(\piPar{\alpha}{Q'})}$} (NuPPrimeQPrime);
  \draw[->] (NuPPrimeQPrime) to node {$\piRestrict{(\piPar{\ren{(\pop{}{0})}{\phi}}{\psi})}$} (NuPDoublePrimeQDoublePrime);
\end{tikzpicture}
}
\end{center}

%% file: fig/residual/illustrate/propagate-1.tex
\begin{center}
\scalebox{\smathparscale}{
\begin{smathpar}
\inferrule*[left={\ruleName{$\piRestrictA{\piInput{x}}{\param}$}}]
{
  \inferrule*[left={\ruleName{$E$}}]
  {
    \vdots
  }
  {
    \suc{\Gamma} \vdash P
    \transitionWithoutSmash{\piInput{\suc{x}}}
    \plus{\Gamma}{2} \vdash R
  }
}
{
  \Gamma \vdash \piRestrict{P}
  \transitionWithoutSmash{\piInput{x}}
  \suc{\Gamma} \vdash \piRestrict{(\ren{\swapR}{R})}
}
\and
\inferrule*[left={\ruleName{$\piRestrictA{\piInput{u}}{\param}$}}]
{
  \inferrule*[left={\ruleName{$E'$}}]
  {
    \vdots
  }
  {
    \suc{\Gamma} \vdash P
    \transitionWithoutSmash{\piInput{\suc{u}}}
    \plus{\Gamma}{2} \vdash R'
  }
}
{
  \Gamma \vdash \piRestrict{P}
  \transitionWithoutSmash{\piInput{u}}
  \suc{\Gamma} \vdash \piRestrict{(\ren{\swapR}{R'})}
}
\end{smathpar}
}
\end{center}

%% file: fig/residual/illustrate/propagate-2.tex
\begin{center}
\scalebox{0.8}{
\begin{tikzpicture}[node distance=1.5cm, auto]
  \node (PQ) [node distance=2cm] {
    $\suc{\Gamma} \vdash P$
  };
  \node (PPushQ) [right of=PQ, above of=PQ] {
    $\plus{\Gamma}{2} \vdash R$
  };
  \node (PushPQ) [below of=PQ, right of=PQ] {
    $\plus{\Gamma}{2} \vdash R'$
  };
  \node (PushPSucPushQ) [node distance=4cm, right of=PPushQ] {
    $\plus{\Gamma}{3} \vdash P'$
  };
  \node (SwapPSwapQ) [below of=PushPSucPushQ] {
    $\plus{\Gamma}{3} \vdash \ren{\swapR}{P'}$
  };
  \node (SucPushPPushQ) [node distance=4cm, right of=PushPQ] {
    $\plus{\Gamma}{3} \vdash P''$
  };
  \draw[->] (PQ) to node [yshift=-1ex] {$\cxtRaw{E}{\piInput{\suc{x}}}$} (PPushQ);
  \draw[->] (PQ) to node [yshift=1ex,swap] {$\cxtRaw{E'}{\piInput{\suc{u}}}$} (PushPQ);
  \draw[->] (PPushQ) to node {$\cxtRaw{(\residual{E'}{E})}{\piInput{\plus{u}{2}}}$} (PushPSucPushQ);
  \draw[->] (PushPQ) to node [swap] {$\cxtRaw{(\residual{E}{E'})}{\piInput{\plus{x}{2}}}$} (SucPushPPushQ);
  \draw[->] (PushPSucPushQ) to node {$\ren{\swapR}{}$} (SwapPSwapQ);
  \draw[->] (SwapPSwapQ) to node {$\phi$} (SucPushPPushQ);
\end{tikzpicture}
}
\end{center}

%% file: fig/residual/illustrate/propagate-3.tex
\begin{center}
\scalebox{\smathparscale}{
\begin{smathpar}
\inferrule*[left={\ruleName{$\piRestrictA{\param}{\param}$}}]
{
  \inferrule*[left={\ruleName{$\ren{\swapR}{\param}$}}]
  {
    \inferrule*[left={\ruleName{$\residual{E'}{E}$}}]
    {
      \vdots
    }
    {
      \plus{\Gamma}{2} \vdash R
      \transitionWithoutSmash{\piInput{\plus{u}{2}}}
      \plus{\Gamma}{3} \vdash P'
    }
  }
  {
    \plus{\Gamma}{2} \vdash \ren{\swapR}{R}
    \transitionWithoutSmash{\piInput{\plus{u}{2}}}
    \plus{\Gamma}{3} \vdash \ren{(\suc{\swapR})}{P'}
  }
}
{
  \suc{\Gamma} \vdash \piRestrict{(\ren{\swapR}{R})}
  \transitionWithoutSmash{\piInput{\suc{u}}}
  \plus{\Gamma}{2} \vdash \piRestrict{(\ren{\swapR}{\ren{(\suc{\swapR})}{P'}})}
}
\end{smathpar}
}
\end{center}

%% file: fig/residual/illustrate/propagate-4.tex
\begin{center}
{\small
\begin{align*}
&\phantom{{}=}\;\ren{\swapR}{\piRestrict{(\ren{\swapR}{\ren{(\suc{\swapR})}{P'}}})}
\\
&= \piRestrict{(\ren{(\suc{\swapR})}{\ren{\swapR}{\ren{(\suc{\swapR})}{P'}}})}
& \text{(definition of $\ren{\param}{}$)}
\\
&= \piRestrict{(\ren{\swapR}{\ren{(\suc{\swapR})}{\ren{\swapR}{P'}}})}
& \text{(\lemref{swap-after-suc-swap-after-swap})}
\\
&\congEq \piRestrict{(\ren{\swapR}{\ren{(\suc{\swapR})}{P''}})}
& \text{($\piRestrict{(\ren{\swapR}{\ren{(\suc{\swapR})}{\phi}})}$)}
\end{align*}
}
\end{center}

%% file: fig/residual/illustrate/extrude-propagate-1.tex
\begin{center}
\scalebox{\smathparscale}{
\begin{smathpar}
\inferrule*[left={\ruleName{$\piParLTau{\param}{\param}{0}$}}]
{
  \inferrule*[left={\ruleName{$\ren{\push{}}{\param}$}}]
  {
    \inferrule*[left={\ruleName{$E$}}]
    {
      \vdots
    }
    {
      \Gamma \vdash P
      \transition{\piInput{x}}
      \suc{\Gamma} \vdash R
    }
  }
  {
    \suc{\Gamma} \vdash \ren{\push{}}{P}
    \transitionWithoutSmash{\piInput{\suc{x}}}
    \plus{\Gamma}{2} \vdash \ren{(\suc{\push{}})}{R}
  }
  \\
  \inferrule*[left={\ruleName{$\residual{F'}{F}$}}]
  {
    \vdots
  }
  {
    \suc{\Gamma} \vdash S
    \transitionWithoutSmash{\piOutput{\suc{x}}{0}}
    \suc{\Gamma} \vdash Q'
  }
}
{
  \suc{\Gamma} \vdash \piPar{\ren{\push{}}{P}}{S}
  \transitionWithoutSmash{\tau}
  \suc{\Gamma} \vdash \piPar{\ren{(\pop{}{0})}{\ren{(\suc{\push{}})}{R}}}{Q'}
}
\end{smathpar}
}
\end{center}

%% file: fig/residual/illustrate/extrude-propagate-2.tex
\begin{center}
\scalebox{\smathparscale}{
\begin{smathpar}
\inferrule*[left={\ruleName{$\piRestrictOutput{\param}$}}]
{
  \inferrule*[left={\ruleName{$\piParR{\piOutput{\suc{u}}{0}}{R}{\param}$}}]
  {
    \inferrule*[left={\ruleName{$\residual{F}{F'}$}}]
    {
      \vdots
    }
    {
      \suc{\Gamma} \vdash S'
      \transitionWithoutSmash{\piOutput{\suc{u}}{0}}
      \suc{\Gamma} \vdash Q''
    }
  }
  {
    \suc{\Gamma} \vdash \piPar{R}{S'}
    \transitionWithoutSmash{\piOutput{\suc{u}}{0}}
    \suc{\Gamma} \vdash \piPar{R}{Q''}
  }
  }
{
   \Gamma \vdash \piRestrict{(\piPar{R}{S'})}
   \transitionWithoutSmash{\piBoundOutput{u}}
   \suc{\Gamma} \vdash \piPar{R}{Q''}
}
\end{smathpar}
}
\end{center}

%% file: fig/residual/illustrate/extrude-propagate-3.tex
\begin{center}
\scalebox{\smathparscale}{
\begin{smathpar}
\inferrule*[left={\ruleName{$\piParLNu{\ren{\push{}}{E}}{\param}$}}]
{
  \inferrule*[left={\ruleName{$\residual{F'}{F}$}}]
  {
    \vdots
  }
  {
    \suc{\Gamma} \vdash S
    \transitionWithoutSmash{\piBoundOutput{\suc{x}}}
    \plus{\Gamma}{2} \vdash Q'
  }
}
{
  \suc{\Gamma} \vdash \piPar{\ren{\push{}}{P}}{S}
  \transitionWithoutSmash{\tau}
  \suc{\Gamma} \vdash \piRestrict{(\piPar{\ren{(\suc{\push{}})}{R}}{Q'})}
}
\end{smathpar}
}
\end{center}

%% file: fig/residual/illustrate/extrude-propagate-4.tex
\begin{center}
\scalebox{\smathparscale}{
\begin{smathpar}
\inferrule*[left={\ruleName{$\piRestrictA{\piBoundOutput{u}}{\param}$}}]
{
  \inferrule*[left={\ruleName{$\piParR{\piBoundOutput{\suc{u}}}{R}{\param}$}}]
  {
    \inferrule*[left={\ruleName{$\residual{F}{F'}$}}]
    {
      \vdots
    }
    {
      \suc{\Gamma} \vdash S'
      \transitionWithoutSmash{\piBoundOutput{\suc{u}}}
      \plus{\Gamma}{2} \vdash Q''
    }
  }
  {
    \suc{\Gamma} \vdash \piPar{R}{S'}
    \transitionWithoutSmash{\piBoundOutput{\suc{u}}}
    \plus{\Gamma}{2} \vdash \piPar{\ren{\push{}}{R}}{Q''}
  }
  }
{
   \Gamma \vdash \piRestrict{(\piPar{R}{S'})}
   \transitionWithoutSmash{\piBoundOutput{u}}
   \suc{\Gamma} \vdash \piRestrict{(\piPar{\ren{\swapR}{\ren{\push{}}{R}}}{\ren{\swapR}{Q''}})}
}
\end{smathpar}
}
\end{center}

%% file: fig/residual/congruence/nu-nu-swap-cases.tex
\begin{figure}[H]
\begin{minipage}[b]{.45\linewidth}
\scalebox{0.8}{
\begin{tikzpicture}[node distance=1.5.8cm, auto]
  \node (1) [node distance=2cm] {$\Gamma \vdash \piRestrict{\piRestrict{(\ren{\swapR}{P})}}$};
  \node (2) [node distance=2cm, below of=1] {$\Gamma \vdash \piRestrict{\piRestrict{P}}$};
  \node (3) [node distance=5.8cm, right of=1] {$\suc{\Gamma} \vdash \piRestrict{R}$};
  \node (4) [node distance=5.8cm, right of=2] {$\suc{\Gamma} \vdash \piRestrict{R}$};
  \draw[->] (1) to node {$\piRestrictOutput{\piRestrictA{\piOutput{\suc{x}}{0}}{E}}$} (3);
  \draw[->] (1) to node [swap] {$\congRestrictSwap{P}$} (2);
  \draw[-,double distance=1pt] (3) to node {} (4);
  \draw[->] (2) to node [swap] {$\piRestrictA{\piBoundOutput{x}}{\piRestrictOutput{(\ren{\swapR}{E})}}$} (4);
\end{tikzpicture}
}
\end{minipage}
\begin{minipage}[b]{.49\linewidth}
\scalebox{0.8}{
\begin{tikzpicture}[node distance=1.5.8cm, auto]
  \node (1) [node distance=2cm] {$\Gamma \vdash \piRestrict{\piRestrict{(\ren{\swapR}{P})}}$};
  \node (2) [node distance=2cm, below of=1] {$\Gamma \vdash \piRestrict{\piRestrict{P}}$};
  \node (3) [node distance=5.8cm, right of=1] {$\Gamma \vdash \piRestrict{\piRestrict{R}}$};
  \node (4) [node distance=5.8cm, right of=2] {$\Gamma \vdash \piRestrict{\piRestrict{(\ren{\swapR}{R})}}$};
  \draw[->] (1) to node {$\piRestrictA{c}{\piRestrictA{c'}{E}}$} (3);
  \draw[->] (1) to node [swap] {$\congRestrictSwap{P}$} (2);
  \draw[->] (3) to node {$\congRestrictSwapInv{R}$} (4);
  \draw[->] (2) to node [swap] {$\piRestrictA{c}{\piRestrictA{c'}{(\ren{\swapR}{E})}}$} (4);
\end{tikzpicture}
}
\end{minipage}
\\
\begin{minipage}[b]{.45\linewidth}
\scalebox{0.8}{
\begin{tikzpicture}[node distance=1.5.8cm, auto]
  \node (1) [node distance=2cm] {$\Gamma \vdash \piRestrict{\piRestrict{(\ren{\swapR}{P})}}$};
  \node (2) [node distance=2cm, below of=1] {$\Gamma \vdash \piRestrict{\piRestrict{P}}$};
  \node (3) [node distance=5.8cm, right of=1] {$\suc{\Gamma} \vdash \piRestrict{(\ren{\swapR}{R})}$};
  \node (4) [node distance=5.8cm, right of=2] {$\suc{\Gamma} \vdash \piRestrict{(\ren{\swapR}{R})}$};
  \draw[->] (1) to node {$\piRestrictA{\piBoundOutput{x}}{\piRestrictOutput{E}}$} (3);
  \draw[->] (1) to node [swap] {$\congRestrictSwap{P}$} (2);
  \draw[-,double distance=1pt] (3) to node {} (4);
  \draw[->] (2) to node [swap] {$\piRestrictOutput{\piRestrictA{\piOutput{x+1}{0}}{(\ren{\swapR}{E})}}$} (4);
\end{tikzpicture}
}
\end{minipage}
\begin{minipage}[b]{.49\linewidth}
\scalebox{0.8}{
\begin{tikzpicture}[node distance=1.5.8cm, auto]
  \node (1) [node distance=2cm] {$\Gamma \vdash \piRestrict{\piRestrict{(\ren{\swapR}{P})}}$};
  \node (2) [node distance=2cm, below of=1] {$\Gamma \vdash \piRestrict{\piRestrict{P}}$};
  \node (3) [node distance=5.8cm, right of=1] {$\suc{\Gamma} \vdash \piRestrict{\piRestrict{(\ren{(\suc{\swapR})}{\ren{\swapR}{R}})}}$};
  \node (4) [node distance=5.8cm, right of=2] {$\suc{\Gamma} \vdash \piRestrict{\piRestrict{(\ren{\swapR}{\ren{(\suc{\swapR})}{\ren{\swapR}{R}}})}}$};
  \draw[->] (1) to node {$\piRestrictA{b}{\piRestrictA{b'}{E}}$} (3);
  \draw[->] (1) to node [swap] {$\congRestrictSwap{P}$} (2);
  \draw[->] (3) to node {$\congRestrictSwapInv{{\ren{(\suc{\swapR})}{\ren{\swapR}{R}}}}$} (4);
  \draw[->] (2) to node [swap] {$\piRestrictA{b}{\piRestrictA{b'}{(\ren{\swapR}{E})}}$} (4);
\end{tikzpicture}
}
\end{minipage}

\crossrule
\caption{Cofinality of $\residual{\phi}{E}$ and $\residual{E}{\phi}$ in
  the $\congRestrictSwap{}$ cases}
\label{fig:residual:congruence:nu-nu-swap-cases}
\end{figure}